\documentclass[lettersize,journal]{IEEEtran}
\usepackage{amsmath,amsfonts}
\usepackage{algorithm}
\usepackage{array}
\usepackage[caption=false,font=normalsize,labelfont=sf,textfont=sf]{subfig}
\usepackage{textcomp}
\usepackage{stfloats}
\usepackage{url}
\usepackage{verbatim}
\usepackage{graphicx}
\usepackage{cite}
\usepackage{etoolbox}
\usepackage{orcidlink}
\usepackage{colortbl}
\usepackage{threeparttable}
\usepackage{booktabs}  
\usepackage{multirow}
\usepackage{tablefootnote}
\usepackage{enumitem}
\usepackage{fontspec}
\usepackage{url}
\usepackage{tablefootnote}
\usepackage{tikz}

\usepackage[algo2e,linesnumbered,ruled,vlined]{algorithm2e}
\usepackage{algpseudocode}

\usepackage[most]{tcolorbox}  
\definecolor{best}{rgb}{.867,  .922,  .969}
\definecolor{second_best}{rgb}{.988,  .894,  .839}

\definecolor{mycolor}{HTML}{BDD7EE}
\newtcolorbox[auto counter, number within=section]{promptbox}[2][]{breakable, colframe=mycolor, colback=mycolor!10!white, coltitle=black, fonttitle=\bfseries, title=Prompt~\thetcbcounter: #2,#1}
\renewcommand{\footnoterule}{%
  \kern -3pt
  \hrule width 0.48\textwidth height 0.4pt
  \kern 2.6pt
}
\usepackage{hyperref}
\begin{document}

\title{$DynamiX$: Large-Scale  Dynamic Social Network Simulator}

\author{Yanhui Sun\orcidlink{0009-0008-0507-7687}, \and Wu Liu\orcidlink{0000-0003-1633-7575},~\IEEEmembership{Senior Member,~IEEE,} \thanks{Y. Sun, W. Liu, W. Wang, H. Yao, and Y. Zhang are with the School of Information Science and Technology, University of Science and Technology of China, Hefei 230022, China. Email:\{liuwu, yaohantao, zhyd73\}@ustc.edu.cn, \{YanhuiS1999, wwt6244\}@mail.ustc.edu.cn.} Wentao Wang\orcidlink{0009-0000-2366-3227}, Hantao Yao\orcidlink{0000-0001-8125-2864},~\IEEEmembership{Member,~IEEE,} Jiebo Luo\orcidlink{0000-0002-4516-9729},~\IEEEmembership{Fellow,~IEEE,} Yongdong Zhang\orcidlink{0000-0002-1151-1792},~\IEEEmembership{Fellow,~IEEE}\thanks{J. Luo is with the Department of Computer Science, University of
Rochester, Rochester, NY 14627, USA. E-mail: jluo@cs.rochester.edu.}}




\markboth{Journal of \LaTeX\ Class Files,~Vol.~14, No.~8, August~2021}%
{Sun \MakeLowercase{\textit{et al.}}: DynamiX: Large-Scale  Dynamic Social Network Simulator}


\maketitle

\begin{abstract}

Social network simulator is a valuable tool for understanding how information spreads and collective behaviors unfold within social platforms. 
The emergence of large language models (LLMs) has enhanced the effectiveness of agent-based simulation frameworks. 
However, existing simulators focus on fixed user roles and static social relationships, limiting their ability to accurately capture real-world social dynamics. 
To address this limitation, we propose a novel large-scale dynamic social network simulator, $DynamiX$, that accounts for the user-role switching and structural evolution of social relationships to better simulate the individual decision-making and collective attitude dynamics. 
Since core agents driven by LLMs play a critical role in information propagation, the Dynamic Hierarchy (DH) module is proposed to identify core agents from ordinary agents for achieving the user-role switching at each timestep. 
Moreover, we propose the distinct modeling strategies of dynamic social relationships for different agent types. 
For core agents, the Personalized Relationships Evolution Engine (PREE) is designed to recommend potential like-minded agents, simulating homogeneous connections and autonomous behaviors decision. 
For ordinary agents, the Dynamic-Relationship-Oriented Agent-Based Model (DRO-ABM) is developed to effectively address unequal social interactions and capture the patterns of relationships adjustments driven by multi-dimensional factors. 
Experimental results demonstrate that $DynamiX$ exhibits marked improvements in attitude dynamics simulation and collective behaviors analysis compared to static simulators. 
Moreover, $DynamiX$ opens up a new perspective on follower growth prediction, providing empirical evidence for cultivating opinion leaders. 

\end{abstract}


\section{Introduction}

\IEEEPARstart{S}{ocial} platforms, as microcosms of real-world society,  have emerged as central mediums for global behaviors interactions, benefiting from extensive connectivity and real-time information exchange \cite{DBLP:journals/eswa/HafieneKR20, JABARILOTF2022126480}. While accelerating the evolution of social dynamics, social platforms also catalyze the spread of misinformation, the polarization of group attitudes, and numerous other negative consequences, such as provoking conflicts, triggering cyber-violence, and even eroding social trust \cite{starbird2014rumors, grinberg2019fake}. In the context, social platforms serve as a natural experimental ground for investigating how information spreads and how collective behaviors unfold, providing insights that are crucial for understanding social progress and maintaining social stability \cite{carr2015social, ruths2014social}.


\begin{figure}[t]
  \centering
  \setlength{\abovecaptionskip}{0.1cm}
  \includegraphics[width=\linewidth]{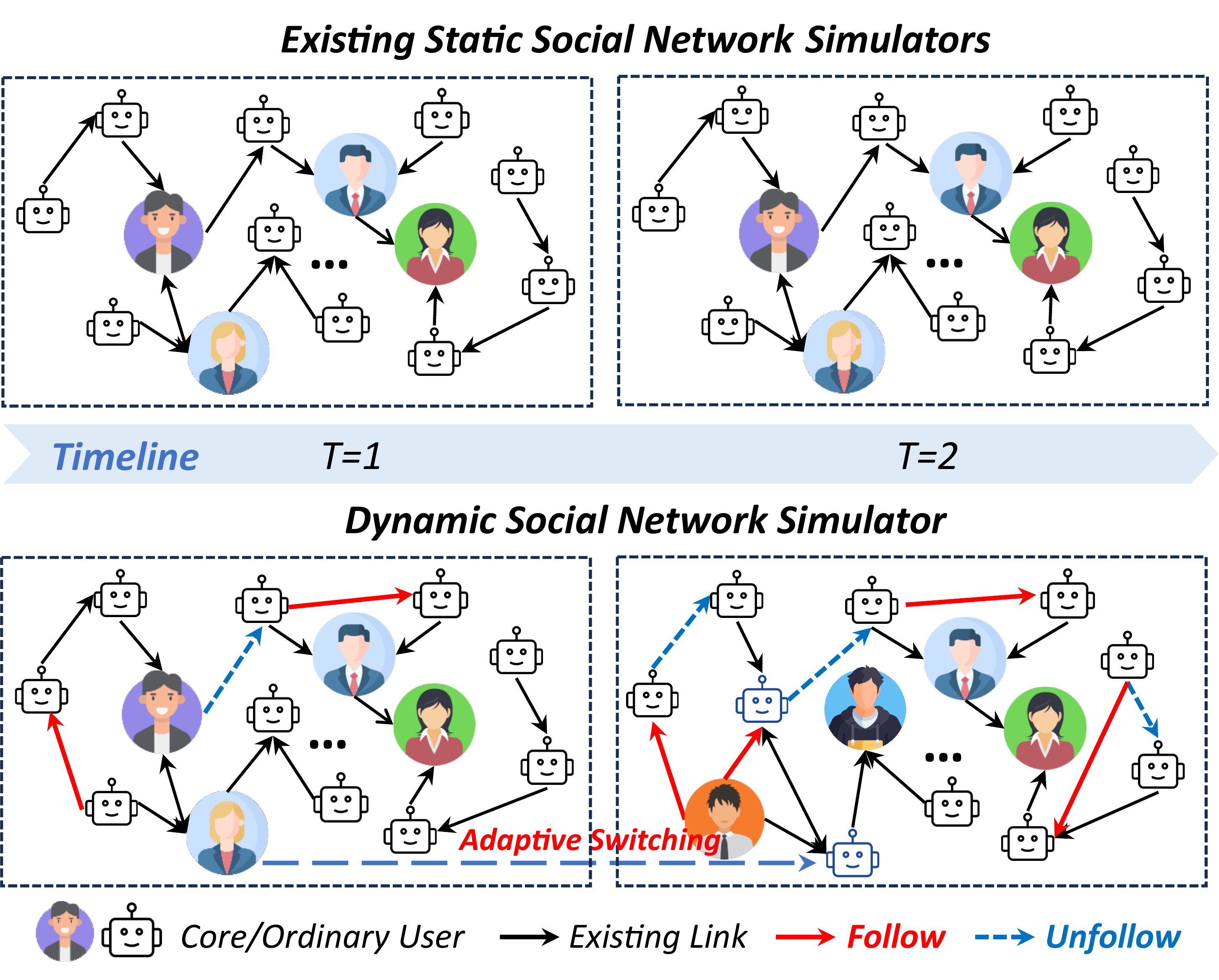}
  \caption{Illustration of simulation results under static versus dynamic social networks. In static networks, social relationships remain unchanged throughout simulations, whereas dynamic networks feature switching core agents and evolving relationships dynamically.}
  \label{Intro}
\end{figure}


Agent-based models (ABMs) are widely utilized to analyze macro-level social dynamics by investigating collective interactions among individuals \cite{DBLP:journals/advcs/DeffuantNAW00,DBLP:journals/jasss/HegselmannK02, DBLP:journals/jasss/DeffuantAWF02, DBLP:journals/cmot/JagerA05, lorenz2021individual}. 
However, these models often struggle to accurately capture micro-level effects of heterogeneous individual behaviors on information propagation, primarily due to their reliance on predefined heuristic rules that insufficiently represent individual behaviors \cite{doi:10.1126/science.1167742,doi:10.1073/pnas.092078899}. 
In contrast, recent advancements in LLM-based social network simulators have demonstrated enhanced individual capabilities in self-awareness, reasoning, and decision-making, allowing for a more accurate reflection of social dynamics \cite{DBLP:journals/corr/abs-2411-11581, DBLP:conf/coling/WangLYC25, DBLP:journals/corr/abs-2412-09237}. 
For instance,  Mou \emph{et al.} introduce a Twitter-like (now X) environment featuring 1,000 agents that effectively simulate actions such as posting and retweeting, successfully replicating both individual decision-making and collective attitude dynamics \cite{DBLP:conf/acl/MouWH24}. 
In social network simulations, the realism of interactions improves as the number of agents increases, thereby enhancing simulation accuracy. 
Thus, scalability has become a central focus, prompting exponential growth in the simulation population \cite{DBLP:journals/corr/abs-2504-10157,DBLP:journals/corr/abs-2502-08691,DBLP:journals/corr/abs-2505-07581}. 
Some simulators escalate the number of agents to over 100,000 by activating a fixed subset of core agents to engage in human-like interactions with a static hierarchical structure \cite{DBLP:journals/corr/abs-2410-04360, DBLP:journals/corr/abs-2411-11581, DBLP:journals/corr/abs-2504-10157}. 
However, reliance on fixed subsets of core agents and static interaction structures fails to account for the continuous switching of user roles  and the dynamic adjustments of social relationships,  as illustrated in Fig. \ref{Intro}.
Consequently, capturing these dynamic characteristics is essential for ensuring that the simulators accurately reflect the complexities of real-world attitude dynamics and social interactions \cite{doi:10.1111/j.1529-1006.2006.00030.x,DBLP:journals/jocss/SasaharaCPCFM21}.

To address this need, an insightful solution is to conduct a dynamic social network simulator.
This endeavor presents two significant challenges: 1) dynamic user roles, and 2) dynamic social relationships. 
Firstly, user roles are diverse and dynamic in large-scale social networks.
Unlike the majority of ordinary roles, core agents driven by LLM naturally exhibit propagation characteristics, such as a higher spread potential \cite{DBLP:conf/kdd/KempeKT03}, making it a nontrivial problem to assign appropriate roles to agents.
Moreover, within dynamic social networks, user roles are not static but continually evolving.
For instance, historical ordinary agents may become core agents for actively engaging in discussions and facilitating the spread of information.
Therefore, modeling the dynamic switching of user roles is a critical component of the social simulator, enhancing decision-making efficiency and accuracy \cite{doi:10.1111/j.1529-1006.2006.00030.x, DBLP:journals/corr/abs-2503-18891}.
Secondly, social relationships form and dissolve frequently in everyday life \cite{statista2024}. 
The dynamic evolution constitutes another essential dimension for understanding information propagation and collective phenomena. 
However, there are several complex factors influencing the formation and dissolution of social relationships. 
For example, users typically prefer to form relationships with like-minded individuals, and reducing interactions with those holding opposing views \cite{nickerson1998confirmation, sears1967selective}. 
Additionally, user influence, unequal interactions, and the quality of content play significant roles in determining the reach and appeal of information, thereby influencing the adjustments of social relationships \cite{DBLP:journals/soco/GirdharMB19, Hangal2010AllFA}. 
Therefore, the intricate interplay among these factors substantially increases the complexity of modeling dynamic social relationships, making it difficult to achieve realistic social network simulations.

In this work, we introduce $DynamiX$, a large-scale social network simulator designed to reflect the dynamic switching of agent roles and capture the structural evolution, \emph{i.e.,} the formation and dissolution, of social relationships.  
Specifically, we design a Dynamic Hierarchy (DH) module to distinguish opinion leaders from ordinary users, by quantifying their propagation potential and content diversity at each timestep. 
This allows adaptive switching of core agents driving event propagation, optimizing the balance between efficiency and accuracy in large-scale social simulations. 
Furthermore, we develop distinct modeling strategies of dynamic social relationships for different agent types. 
For LLM-based core agents, we model the evolution of social relationships as continuous responses, \emph{e.g.,} follow and unfollow, to incoming information streams. 
We propose a Personalized Relationships Evolution Engine (PREE) to recommend like-minded information streams from non-neighbors, promoting form a new follow relationship. 
Based on interacting memories from neighbors and personalized information streams, core agents autonomously decide whether to unfollow or follow others, thereby mimicking realistic homogeneous connection behaviors and dynamic relationships adjustments. 
For ABM-driven ordinary agents, we construct a Dynamic-Relationship-Oriented Agent-Based Model (DRO-ABM). 
It introduces a concept of trust to quantify unequal interactions among agents and employs a multi-factor relationships predictor to model relationships change driven by multi-dimensional factors. 
This strategy reflects how core agents and local neighbors influence the passive behaviors of most agents in the real world, while maintaining scalability for large-scale simulations.

Experiments conducted on real-world event propagation datasets show that $DynamiX$ achieves marked improvements in predicting attitude dynamics and reproducing collective behaviors phenomena. 
Compared to static social networks, $DynamiX$ effectively models the dynamic evolution of public attitudes with greater stability and adaptability.
It also accelerates the emergence of attitude polarization, with new follow relationships displaying a noticeable clustering effect. 
Furthermore, $DynamiX$ offers valuable insight for social network simulators by predicting follower growth during event propagation. 
The experimental findings reveal that high-influence users can significantly increase their follower counts through trending promotion, while low-influence users necessitate sustained high-quality content, providing empirical evidence for cultivating opinion leaders.

In summary, our contributions are threefold:
\begin{itemize}[leftmargin=1em]
\item We introduce $DynamiX$, a large-scale social network simulator, expressly designed for modeling dynamic social networks, providing a high-fidelity experimental platform for simulating social dynamics, reproducing  collective phenomena, and predicting follower growth.
\item We develop a dynamic hierarchy module with agent roles assessment,  identifying core agents with key characteristics from the majority of ordinary agents, allowing the adaptive switching of core agents and efficient large-scale simulations.
\item We develop distinct strategies for modeling dynamic social relationships across agent types. Core agents achieve behaviors decision through personalized relationships recommendations, while ordinary agents rely on a multi-factor relationships predictor, resulting in a more faithful and fine-grained simulation of social dynamics.

\end{itemize}

\section{Related Work}


\subsection{LLM-based Social Network Simulators}

Social science seeks to understand human behaviors within societal contexts, offering  critical insights into how societies function and evolve. Traditional methods, such as  interviews\cite{knottInterviewsSocialSciences2022a}, and controlled experiments \cite{doi:10.1073/pnas.2014893118}, have long been instrumental in exploring social phenomena. However, they often encounter challenges of high costs, ethical constraints, scalability, and replication. To this end, agent-based models (ABMs) have emerged as a computational alternative \cite{DBLP:journals/advcs/DeffuantNAW00,DBLP:journals/jasss/HegselmannK02,DBLP:journals/jasss/DeffuantAWF02,DBLP:journals/cmot/JagerA05,lorenz2021individual}, allowing for in silico experimentation on social dynamics by flexibly simulating interactions through predefined rules. Yet ABMs often rely on heuristic algorithms and simplified behaviors, limiting their ability to capture the nuances of human cognition and real-world social interactions. Leveraging the strong capabilities of LLMs in simulating complex individual behaviors, such as maintaining personality traits, exhibiting self-awareness, and expressing diverse emotions, recent studies have shown that LLM-based social network simulators open up  new prospects for social simulation\cite{wang2023user,DBLP:conf/naacl/ChuangGHSHYSHR24}. Chronologically, LLM-based simulators can be divided into two distinct  research stages.

\textbf{Complex Interactions:} In the first stage, research efforts primarily focus on modeling interaction mechanisms tailored to specific scenarios, typically involving fewer than 1,000 agents. For example, Generative Agents simulates the daily interactions among 25 agents in a virtual town,  demonstrating that LLM-based agents can generate social behaviors indistinguishable from human-created content \cite{DBLP:conf/uist/ParkOCMLB23}. FPS focuses on modeling the dissemination of fake news within small communities, providing a detailed analysis of the propagation trends and intervention  mechanisms \cite{DBLP:conf/ijcai/LiuCZG0024}. TrendSim simulates the impact of poisoning attacks on trending topics in social media \cite{zhang-etal-2025-trendsim}. Wang \textit{et al.} simulate social interactions within classical network structures, validating its effectiveness in evaluating and countering polarization phenomena \cite{DBLP:conf/coling/WangLYC25}.

\textbf{Large-scale Simulations:} In the second stage, small-scale simulations are found inadequate for capturing the complexity and generalizability required in social science, prompting the development of scalable, general-purpose social simulators to support large agent populations. AgentSociety constructs an urban simulator with a realistic societal environment where over 10k agents emulate diverse social phenomena, including polarization, messages spread, economic effects, and external shocks \cite{DBLP:journals/corr/abs-2502-08691}. GenSim provides a versatile simulation platform that supports modular functions for scenario customization, enabling large-scale simulations involving up to 100,000 agents \cite{DBLP:journals/corr/abs-2410-04360}. SocioVerse features four powerful alignment components and a large real-world user pool to achieve accurate simulations of large-scale agents on social, political, and economic topics \cite{DBLP:journals/corr/abs-2504-10157}. 


Despite advancing the intelligence and scalability of social simulations, most rely on static social networks and overlook the fact that the roles of core agents who drive event propagation also evolve continuously. In this work, we introduce a large-scale social network simulator that reflects the evolving roles of core agents and captures how users dynamically adjust their social relationships.

\subsection{Link Prediction}

Link prediction is a core technique for modeling dynamic social relationships, providing theoretical foundations for uncovering the evolution mechanisms of social relationships, and the patterns of individual behaviors. They can be broadly categorized into three types: 


\textbf{Heuristic Methods:} They typically leverage network topology to assess structure similarity between node pairs \cite{DBLP:conf/cikm/Liben-NowellK03,DBLP:conf/socialcom/FireTLPRE11,DBLP:journals/socnet/AdamicA03,DBLP:journals/jasis/Liben-NowellK07,chen2012identifying}. For example, Michael \textit{et al.} propose a set of simple structure features to identify missing links, thereby uncovering hidden relationships within social networks \cite{DBLP:conf/socialcom/FireTLPRE11}. While these methods offer advantages of easy implementation and low computation cost, they are inherently limited by their reliance on static structure information. As they ignore temporal variations and attribute similarity, they struggle to capture the dynamic and heterogeneous nature of social networks, thereby limiting their effectiveness in modeling link evolution.

\textbf{Probabilistic Methods:} To address the limitations, researchers have introduced probabilistic methods, which construct parameterized statistical models to simulate the links formation through estimating the underlying connection probabilities between nodes \cite{DBLP:conf/icdm/DongTWTCRC12, DBLP:conf/icc/DasD17}. For instance, RFG captures key network evolution mechanisms by leveraging common social patterns and structure features across heterogeneous networks \cite{DBLP:conf/icdm/DongTWTCRC12}. Despite their theoretical rigor, such methods often suffer from high data dependency, complex model design, and strong prior assumptions, which constrain their scalability and adaptability in dynamic and evolving environments.

\textbf{Graph-based Deep Learning Methods:} Recently, GNN have been widely adopted for link prediction, as they facilitate the automatic learning of latent node representations to better capture complex structural and semantic patterns \cite{DBLP:journals/pami/YinSXGCHLL24,pan2016predicting,DBLP:journals/pami/CaiLWJ22,DBLP:journals/pami/FanZWLZGD22}. LGLP transforms original graphs into line graphs to explicitly model link relations, achieving superior performance on sparse and structure-sensitive networks \cite{DBLP:journals/pami/CaiLWJ22}. HeteHG-VAE transforms multi-level dependencies into hypergraphs and learns heterogeneous information of nodes and hyperedges through a bayesian generative framework, effectively capturing pairwise and high-order semantic relations \cite{DBLP:journals/pami/FanZWLZGD22}.

While recent methods have advanced performance, they predominantly focus on structural information, overlooking the multi-dimensional social factors in relationships evolution. This limitation hinders interpretability and generalization in long-term high-fidelity social simulations. To this end, we propose novel link prediction methods tailored for social network simulation, integrating multi-dimensional factors including user persona, attitude similarity, and unequal social interactions, to enhance the interpretability and accuracy of  large-scale simulations.

\begin{figure*}[t]
  \centering
  \setlength{\abovecaptionskip}{0.1cm}
  \includegraphics[width=0.955\linewidth]{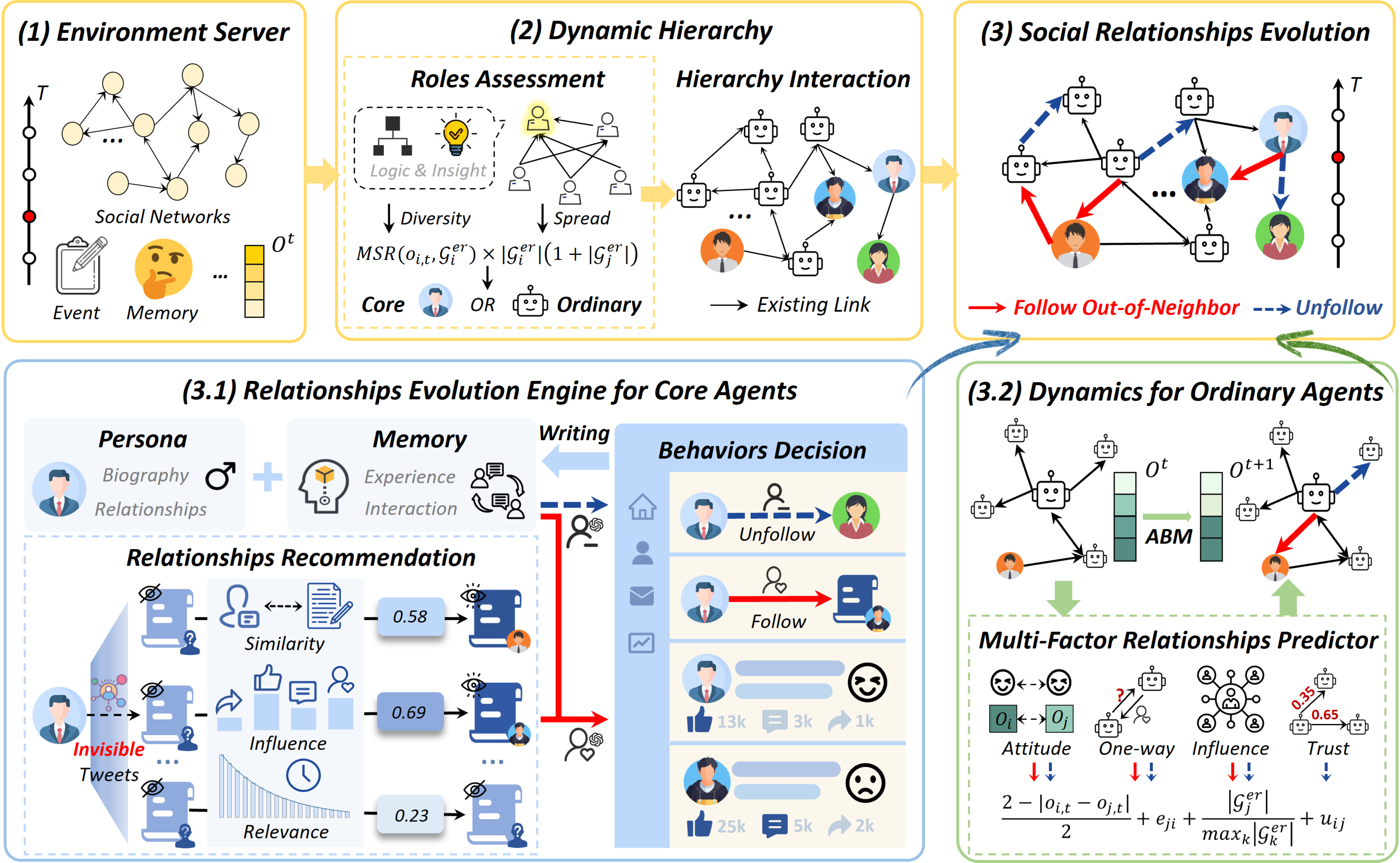}
  \caption{The  architecture of $DynamiX$ framework. Upon receiving a simulation query, the \textit{Environment Server} is initialized. At each timestep, the \textit{Dynamic Hierarchy} partitions users into Core Agents and Ordinary Agents. \textit{Core Agents} autonomously decide their interaction behaviors and adjust social relationships based on the interaction memories received from their followees and recommended non-neighbor agents, whereas \textit{Ordinary Agents} update attitudes and evolve social ties through a dynamic-relationship-oriented agent-based model. This process is iteratively executed to ultimately simulate attitude dynamics, social effects, and follower growth prediction.}
  
  \label{fig:framework}
\end{figure*}

\section{Method}

\subsection{Simulation Framework}  

Social network is modeled as a directed graph $\mathcal{G} = (\mathcal{A}, \mathcal{E})$, with $\mathcal{A} = \{a_1, a_2, \cdots, a_N \}$ delineating the agent population, and $e_{ij} \in \mathcal{E}$ indicating that agent $a_i$ follows agent $a_j$. 
At timestep $t$, agent $a_i$ receives message $m_{j,t}$ from the agent $a_j$ belonging to its follower list $\mathcal{G}_i^{er}$ = $\{a_j | e_{ji} \in \mathcal{E}  \}$ and followee list $\mathcal{G}_i^{ing}$ =  $\{a_j | e_{ij} \in \mathcal{E} \}$ to update its attitude $o_{i,t} \in [-1, 1]$. 
By considering the generated attitudes of all agents within several timesteps, the simulator aims to align its attitude dynamics and collective behaviors with the real world.

To balance simulation accuracy and efficiency, some simulators adopt a static hierarchical design in which LLM-based core agents coexist with ABM-driven ordinary agents \cite{DBLP:conf/acl/MouWH24,DBLP:journals/corr/abs-2411-11581, DBLP:journals/corr/abs-2504-10157}. 
However, such static structures cannot accommodate the inherently dynamic characteristics of social networks, including the continuous switching of user roles and the structural adjustments of social relationships. 
To address this gap, we propose a large-scale dynamic social network simulator named \emph{DynamiX}.
At each timestep, an \textbf{Environment Server} (Fig. \ref{fig:framework} (1)) initializes the social network simulation environment. 
Then, the \textbf{Dynamic Hierarchy} module (Fig. \ref{fig:framework} (2)) considers the key characteristics driving event propagation to dynamically switch agent roles, dividing $\mathcal{A}$ into core agents $\mathcal{A}_{t}^{c}$ and ordinary agents $\mathcal{A}_{t}^{o}$.
And a dynamic social relationships prediction strategy (Fig. \ref{fig:framework} (3)) is proposed to model how agent $a_i$, with different roles, adjusts its social relationships by following or unfollowing agent \( a_j \).
Specifically, \textbf{Core Agents} are equipped with Personalized Relationships Evolution Engine (PREE) to capture  decision-making processes with heterogeneous individual design (Fig. \ref{fig:framework} (3.1)). 
Then, a personalized relationships recommendation method is proposed to present tweets from like-minded non-neighbors, promoting homogeneous connections. 
Based on the above, core agent finally determines whether to adjust the social relationships through the follow-unfollow autonomous behaviors decision.
For \textbf{Ordinary Agents} (Fig. \ref{fig:framework} (3.2)), we utilize Dynamic-Relationship-Oriented Agent-Based Model (DRO-ABM) to numerically simulate attitude dynamics and to achieve the adaptive evolution of social relationships through its multi-factor relationships predictor.
Consequently, the following list \( \mathcal{G}^{ing}_{i} \) of agent \( a_i \) and follower list \( \mathcal{G}^{er}_{j} \) of agent \( a_j \) adaptively evolve over time.

After iterative simulations, $DynamiX$ facilitates efficient large-scale simulations and accurate alignment with real-world social dynamics, revealing how dynamic social networks influence the evolution of public attitudes and collective behaviors, with further details provided in Algorithm \ref{algorithm_lb}. 

\subsection{Dynamic Hierarchy with Agent Roles Assessment}
\label{Dynamic Hierarchy}

When scaling up agent populations, existing simulators typically employ random or static strategies to select a small subset of core agents, equipped with LLM-based interactions, thereby improving computational efficiency \cite{DBLP:conf/acl/MouWH24, DBLP:journals/corr/abs-2411-11581}. 
Nevertheless, such strategies fail to capture continuous switching of core agents across different timesteps, compromising simulation accuracy when the populations of core agents remain finite \cite{doi:10.1111/j.1529-1006.2006.00030.x,DBLP:journals/corr/abs-2503-18891}. 
To address this issue, we propose a Dynamic Hierarchy (DH) module to adaptively identify core agents from ordinary agents across different timesteps, supporting the adaptive switching of core agent roles.

Unlike ordinary roles, core agents naturally exhibit remarkable characteristics, such as spread capability \cite{DBLP:conf/kdd/KempeKT03, kitsakIdentificationInfluentialSpreaders2010} and content diversity \cite{doi:10.1126/science.aaa1160}, both of which play crucial roles in attitude dynamics within social networks. The former characterizes the depth and breadth of information propagation, while the latter provides diverse and broad knowledge to mitigate the formation of echo chambers \cite{DBLP:journals/pnas/CinelliMGQS21}. Based on their significance, we develop a comprehensive influence metric $\phi_{i,t}$ to identify core agents with high spread potential $s_{i,t}$ and content diversity $d_{i,t}$ at each stage of propagation,
\begin{equation}
  \phi_{i,t} = s_{i,t} \times d_{i,t}.
\label{eq:1}
\end{equation}

The metric $s_{i,t}$ is tailored for social networks with follow relationships. 
We employ follower counts rather than followee counts, as information typically propagates from followed agents to their followers. 
To mitigate computational complexity in large-scale networks, the spread potential $s_{i,t}$ of agent $a_i$ in Eq.~\eqref{eq:1} is formulated as second-order follower counts by jointly counting its direct followers $\mathcal{G}_i^{er}$, along with their corresponding followers $\sum_{j\in \mathcal{G}_i^{er}}\sum_{k\in \mathcal{G}_j^{er}}e_{kj}e_{ji}$,
\begin{equation}
  s_{i,t} = \sum\nolimits_{j\in \mathcal{G}_i^{er}}e_{ji} + \sum\nolimits_{j\in \mathcal{G}_i^{er}}\sum\nolimits_{k\in \mathcal{G}_j^{er}}e_{kj}e_{ji}.
\end{equation}

Moreover, to avoid reinforcing similar attitudes among their followers, we formulate the content diversity $d_{i,t}$ in Eq.~\eqref{eq:1} as the attitude variance between $o_{i,t}$ of agent $a_i$ and $o_{j,t}$ of all its followers.
Concurrently, the heterogeneity of followers attitudes enhances the content diversity of core agents, fostering a more dynamic exchange of opinion,
\begin{equation}
  d_{i,t} = \sqrt{\frac{1}{|\mathcal{G}_i^{er}|}\sum\nolimits_{j\in \mathcal{G}_i^{er}}(o_{i,t} - o_{j,t})^2}.
\end{equation}

Once obtaining the influence score of whole agents at each timestep, \emph{i.e.,} $\Phi=\{\phi_{1,t},\phi_{2,t},...,\phi_{N,t}\}$, the top-$k$ agents with the highest $\phi_{i,t}$ treated as core agents $\mathcal{A}^{c}_t$, while the remaining agents are designated as ordinary agents $\mathcal{A}^{o}_t$, forming a dynamic hierarchy that evolves as agent roles switch adaptively,
\begin{equation}
  \mathcal{A}^{c}_t = \{a_{i} \in \mathcal{A}\, | \, i \in \text{TopK}(\Phi) \}, \quad \mathcal{A}^{o}_t = \mathcal{A} \setminus \mathcal{A}^{c}_t.
\end{equation}

One critical aspect for simulating the dynamic social network is that the core agents and ordinary agents have different types of interactions, \emph{i.e.,} core agents interact via language dialogues  (Sec. \ref{core_agent}), while ordinary agents use ABMs for message transmission (Sec. \ref{ordinary_agent}). 
The content generated by core agents can also guide the attitude updates of ordinary agents through ABMs. 
Through iterative simulations, the dynamic hierarchy module can distinguish core agents with key characteristics, achieving the adaptive switching of core agent roles within evolving networks. 
This facilitates accurate alignment with real-world attitude dynamics and supports efficient large-scale social simulation.

\subsection{Personalized Relationships Evolution Engine}
\label{core_agent}

Since LLM-based core agents serve as opinion leaders, we propose a 
Personalized Relationships Evolution Engine (PREE) to enhance their robust personalization and powerful expressive capabilities.
We first equip core agents with persona and memory modules to capture the inherent heterogeneity in individual decision-making. 
After that, to facilitate the formation and dissolution of social relationships, a Personalized Relationships Recommendation is proposed to uncover like-minded tweets from their non-neighbors, promoting homogeneous connection behaviors. 
Finally, by utilizing personalized information streams and messages from neighbors, core agents achieve Autonomous Behaviors Decision-Making by assessing whether the viewpoint of a given agent  is congruent with their own mental state and memories.


\textbf{Heterogeneous Individual Design:} 
To enhance the quality of heterogeneous decision-making and facilitate the change of social relationships, we equip core agents with persona and memory modules to model individual attributes and reflect on past interactions.
Based on demographic distribution \cite{DBLP:conf/hicss/BrunkerWMM20}, we construct the persona $P_i$ of agent $a_i$ with attributes closely associated with attitude stance and behaviors decision, \emph{e.g.,} name, age, gender, occupation, interest, and personality traits. 
Since both historical behaviors of the agent itself and historical interactions from its visible neighbors impact the decision-making accuracy, 
we incorporate two types of memories to represent them: personal experience $\boldsymbol{M}_{i}^{P}$ and environment interaction $\boldsymbol{M}_{i}^{E}$. 
Before executing behaviors decision, the agent employs a retrieval function $\mathcal{F}_{r}(\cdot)$ to retrieve the most relevant memories $\boldsymbol{M}_{i,t}^{R}$:
\begin{equation}
\boldsymbol{M}_{i,t}^{R} =\mathcal{F}_{r}(\mathcal{F}_p({env}_{d}, P_i),\{\boldsymbol{M}_i^P, \boldsymbol{M}_i^E\}),
\end{equation}
where $\mathcal{F}_p(\cdot)$ generate queries for retrieving memories related to the event description $env$ and persona $P_i$ of agent $a_i$.
More details about the constructed persona and memories are provided in the Appendix. 


\textbf{Personalized Relationships Recommendation:} 
To model the formation and dissolution of social relationships, the dynamic nature of social relationships is formulated as reactions, \emph{e.g.,} follow and unfollow, to incoming information streams. \cite{ DBLP:conf/www/YangLSSZZ11, nickerson1998confirmation, sears1967selective}. Unlike the dissolution of social relationships, where core agents assess whether the information streams received from neighbors are consistent with their own position, establishing new relationships requires exposure to information streams from non-neighbors, as depicted in Fig. \ref{fig:framework} (3.1). 
Based on the streams,  core agents then choose to follow those like-minded agents whose views align with their psychological states and memories.


To capture this mechanism, we propose a Personalized Relationships Recommendation (PRR) $\mathcal{F}_{rec}$ to measure the match between agent $a_i$ and candidate tweet $\boldsymbol{L}_j$ with a recommendation score $s_{ij}^{rec}$,
\begin{equation}
      s^{rec}_{ij}= \mathcal{F}_{rec}(a_i,\, \boldsymbol{L}_{j}, \mathcal{N}_i),
\end{equation}
where $\boldsymbol{L}_j$ denotes the $j$-indexed generated tweet in the tweet list $\boldsymbol{L}$, which includes the author persona, tweet content, posting time, and  content attractiveness. To reduce computational costs in large-scale simulations, we retain only candidate tweets $\boldsymbol{L}_{\mathcal{N}_i}$ = $\{\boldsymbol{L}_j\,| \,author(\boldsymbol{L}_j) \in\mathcal{N}_i\}$ generated from the agent set $\mathcal{N}_i$, which consists of followees-of-followees and high-follower non-neighbor agents with similar stances. 
After acquiring the recommendation scores for these tweets at timestep $t$,   \emph{i.e.,} $\boldsymbol{s}_{i}^{rec} = \{{s}_{i1}^{rec},{s}_{i2}^{rec},...,{s}^{rec}_{i|\boldsymbol{L}_{\mathcal{N}_i}|}\}$,
the top-$K$ tweets with the highest $s_{ij}^{rec}$ scores are selected to form the personalized information stream $\boldsymbol{R}_{i,t}$, thereby facilitating the formation of new social relationships,
\begin{equation}
\boldsymbol{R}_{i,t} = \{\boldsymbol{L}_j\, |\, j =  \text{TopK}(\boldsymbol{s}_{i}^{rec})\}.       
\end{equation}

A notable point is that only well-matched information streams from non-neighbors can effectively represent these potential followees.
Since factors that influence the formation of personalized information streams are complex, the $s^{rec}_{ij}$ recommends potential like-minded tweets by quantifying attitude similarity, content relevance, and tweet influence.
To be more precise, it integrates attitude similarity $s_{\text{match}}$ between the agent embedding $\boldsymbol{u}_i \in \mathbb{R}^d$ and tweet embedding $\boldsymbol{p}_j \in \mathbb{R}^d$. 
Moreover, $s_{ij}^{rec}$ includes the temporal relevance of the candidate tweet $\boldsymbol{L}_{j}$ and its content influence $\tau_j$, as formalized in the following,
\begin{equation}
  s_{ij}^{rec} = \underbrace{\cos(\boldsymbol{u}_i,\boldsymbol{p}_j)}_{\text{content matching}} \times \underbrace{e^{-\beta(t-t_{j})}}_{\text{relevance}} \times \underbrace{(1+\tau_j)}_{\text{content influence}},
  \label{eq:score}
\end{equation}
where the temporal relevance follows an exponential decay with rate $\beta$, and $t_{j}$ denotes the post time. $\tau_j$ quantifies the inherent content attractiveness of $\boldsymbol{L}_{j}$ by weighting the number of likes, retweets, comments, and followers.

The method allows core agents to receive information streams outside their social relationships, which align with their persona, sentiment, and responses.
Notably, while the design of the score $s_{ij}^{rec}$ may vary across social platforms, our focus is on utilizing personalized information streams to facilitate autonomous decision-making regarding whether to adjust their social relationships.

\textbf{Autonomous Behaviors Decision-Making:} The behaviors of core
agents reflect their stances towards specific events, encompassing action set $\mathcal{S}_{action}$: \textit{Follow}, \textit{Unfollow}, \textit{Post}, \textit{Retweet}, \textit{Reply},  \textit{Like}, and \textit{Doing Nothing}. Each action is intricately tied to maintaining persona, exhibiting self-awareness, and expressing context-sensitive emotions.
Based on the persona $P_i$, the retrieval memories $\boldsymbol{M}^{R}_{i,t}$, and personalized information streams $\boldsymbol{R}_{i,t}$, the decision-making process $\mathcal{F}_{dm}$  determines how the agent $a_i$ captures the diversity in interacting with neighbors autonomously,
\begin{equation}
    \mathcal{S}_{i,t} = \mathcal{F}_{dm}(P_i, \boldsymbol{M}_{i,t}^{R}, \boldsymbol{R}_{i,t}, env, \mathcal{S}_{action}),
\end{equation}
where $\mathcal{S}_{i,t}$ represents the behaviors set executed at time $t$, selected from the action set $\mathcal{S}_{action}$. 
In particular, core agent $a_i$ evaluates whether the stance of the information stream from their neighbors aligns with their persona and memories, thereby autonomously deciding whether to sever existing relationship with agent $a_j$.
Given the personalized information streams, core agent $a_i$ follows like-minded agent $a_k$ whose tweet satisfies their inner needs, \emph{e.g.,} recognition, curiosity. Consequently, the social networks adaptively evolve over time,
\begin{equation}
  \begin{aligned}
      \scalebox{0.9}{\textit{Unfollow}}\,(a_i,  a_j):  \mathcal{G}_i^{ing} = \mathcal{G}_i^{ing}  \setminus  \{a_j\}, \,\mathcal{G}_j^{er} = \mathcal{G}_j^{er}  \setminus  \{a_i\},\\
    \scalebox{0.9}{\textit{Follow}}\,(a_i, a_k):
    \mathcal{G}_i^{ing} = \mathcal{G}_i^{ing} \cup \{a_k\}, \,\mathcal{G}_k^{er} = \mathcal{G}_k^{er}  \cup  \{a_i\}. 
  \end{aligned}
  \end{equation}

Meanwhile, the tweet $\boldsymbol{L}_{i}$, generated from the behaviors $\mathcal{S}_{i,t}$, provides the attitude $o_{i,t}$ of agent $a_i$ at the timestep $t$. This will guide the attitude updates of ordinary agents via ABMs. 
Additionally, the tweet serves to update the personal experience $\boldsymbol{M}^{P}_{i}$ of agent $a_i$ and environment interaction memories $\boldsymbol{M}^{E}_{j}$ of its followers $a_j \in \mathcal{G}_i^{er}$, thereby influencing the subsequent behavioral decisions, 
\begin{equation}
  o_{i,t} = \mathcal{F}_{score}(\boldsymbol{L}_{i}, env),
\end{equation}
\begin{equation}
  \boldsymbol{M}^{P}_{i} = \boldsymbol{M}^{P}_{i} \cup \{\boldsymbol{L}_{i}\}, \,\,\, \boldsymbol{M}^{E}_{j} = \boldsymbol{M}^{E}_{j} \cup \{\boldsymbol{L}_{i}\},
\end{equation}
where the detailed prompt for $\mathcal{F}_{score}(\cdot)$ is documented in the Appendix.

Through iterative interactions, core agents adapt to the dynamic evolution of social relationships,  driven by messages received from their neighbors and personalized information streams. This continuous adjustments of social relationships emphasizes the role of dynamic social networks in shaping attitudes dynamics and collective behaviors, thus providing a richer and more realistic representation of societal evolution.

\begin{algorithm*}[htbp]

  \caption{$DynamiX$: Large-Scale Dynamic  Social Network Simulator}
  \label{algorithm_lb}
  \KwIn{Environment context description $env$, core agents number $k$, simulation timesteps $T$, and agents set $\mathcal{A} = \{a_1, ..., a_N\}$ with their personas $\{P_i\}_{i=1}^N$,  follower list $\{\mathcal{G}_i^{er}\}_{i=1}^N$, following list $\{\mathcal{G}_i^{ing}\}_{i=1}^N$,  initial attitude $\{o_{i,0}\}_{i=1}^N$}

  \Begin{
      
      \For {each timestep $t$ in 1 to $T$} 
      {
          Calculate influence metric $\phi_{i,t}$ for each agent;

          Group $\mathcal{A}$ into core agents $\mathcal{A}^{c}_t$ and ordinary agents  $\mathcal{A}^{o}_t$ based on the top-$k$ highest values of $\phi_{i,t}$.

          \For{each agent $a_i$ in $\mathcal{A}^{c}_t$}
          {

              Retrieve the most relevant memories $\boldsymbol{M}^{R}_{i,t}= \{\boldsymbol{M}^P_{i}, \boldsymbol{M}_i^E\}$ based on relevance, importance, and timeliness;

              Use personalized relationships recommendation method to present like-minded tweets $\boldsymbol{R}_{i,t}$;

              Generate behaviors set $\mathcal{S}_{i,t}$  based on $env$, $\boldsymbol{R}_{i,t}$, $\boldsymbol{M}_{i,t}^{R}$, and $P_i$;

              Calculate  attitude score $o_{i,t}$ according $\mathcal{S}_{i,t}$;

              Update its personal experience $\boldsymbol{M}_i^P$ and the experiment interaction memories $\boldsymbol{M}^E_j$ of its followers.
          }

          The attitudes of core agents affect the attitude updates of ordinary agents through ABMs.

          \For{each agent $a_i$ in $\mathcal{A}^{o}_t$}
          {
              Employ $\mathcal{F}_{select}$ to determine the agents set $\mathcal{J}_{i,t}$ to interact with;

              Update $o_{i,t}$ by $\mathcal{F}_{update}$ in dynamic-relationship-oriented agent-based model;

              Transmit information to followers through $\mathcal{F}_{message}$;

              \If{$(t -T_{start} )\% T_{interval} = 0$}
                      {
                          Perform the multi-factor relationships predictor $\mathcal{F}_{predict}$ with $p_{follow}$ and $p_{unollow}$.
                      }
          }
          Modular content update of the Environment Server.
      }
  }
  \KwOut{Core agents $\mathcal{A}^{c}_t$, tweet list $\boldsymbol{L}$, attitude $\{o_{i,t}\}_{i=1}^{N}$, follower list $\mathcal{G}_i^{er}$ and following list $\mathcal{G}_i^{ing}$ at each timestep.}
\end{algorithm*}

\subsection{Dynamic-Relationship-Oriented Agent-Based Model}
\label{ordinary_agent}

To balance simulation efficiency and scale, some simulators adopt ABMs to describe interactions among ordinary agents \cite{DBLP:conf/acl/MouWH24,DBLP:journals/corr/abs-2504-10157}. 
However, traditional ABMs face limitations when applied to complex social environments. Firstly, social relationships remain heterogeneous, with different followees exerting varying influence on the attitude update \cite{DBLP:journals/soco/GirdharMB19}, whereas the homogeneous assumption inherent in ABMs fails to realistically capture such unequal interactions. Secondly, ABMs typically lack mechanisms for modeling dynamic social relationships, limiting their ability to respond to event propagation accurately. 
By extending the traditional ABMs, we propose a Dynamic-Relationship-Oriented Agent-Based Model (DRO-ABM).
It introduces a concept of trust $u_{ij} \in [0, 1]$ to quantify unequal interactions among agents and models the relationships adjustments through a multi-factor relationships predictor.
More precisely, it integrates the selection $\mathcal{F}_{select}$, update $\mathcal{F}_{update}$, message $\mathcal{F}_{message}$, and relationships predictor $\mathcal{F}_{predict}$ functions to model distinct agent behaviors, as described below.

\textbf{Selection Function} $\mathcal{F}_{select}$ aims to select the followee set that influences the attitude update of agent $a_i$ at timestep $t$. 
The followee set $\mathcal{J}_{i,t}$ contains agents from the following list $\mathcal{G}_i^{ing}$, whose attitudes are close to $a_i$ with a similarity threshold $\epsilon$. 
Moreover, we equip each agent with a trust boundary $\hat{u}_i$ to mimic real-world unequal interactions, and the agents whose trust level $u_{ij}$ exceeds $\hat{u}_i$ are considered as selected neighbors $\mathcal{J}_{i,t}$,
\begin{equation}
  \mathcal{F}_{select}: \, \mathcal{J}_{i,t} 
  = \{ j |(j \in \mathcal{G}_i^{ing}) \land (|o_{j,t} - o_{i,t}| < \epsilon)  \land (u_{ij} \geq  \hat{u}_i) \}.
\end{equation}
%

\textbf{Update Function} $\mathcal{F}_{update}$ updates the attitude $o_{i,t}$ of agent $a_i$ as a weighted combination of its current attitude and the messages received from its selected followees $\mathcal{J}_{i,t}$, with the weight parameter $\alpha$,
\begin{equation}
  \mathcal{F}_{update}: \, o_{i,t} = \alpha o_{i,t-1} + (1-\alpha){\sum\nolimits_{j \in \mathcal{J}_{i,t}} \omega_{ij} m_{j,t}},
    \label{eq6}
\end{equation}
where $\omega_{ij}$ represents the influence weight of each message $m_{j,t}$ from its followee $a_j$ has on the attitude of agent $a_i$. 
This serves as a direct reflection of unequal interactions. 
Users are more likely to be influenced by friends with higher trust or greater influence.
We thus aim to assign higher weights $\omega_{ij}$ to agents with higher trust or more followers,
\begin{equation}
  \omega_{ij}=\frac{\lambda u_{ij}}{\sum_{k\in \mathcal{J}_{i,t}}u_{ik}}+\frac{(1-\lambda){|\mathcal{G}_j^{er}|}}{\sum_{k \in \mathcal{J}_{i,t}} |\mathcal{G}_k^{er}|},
  \label{eq9}
\end{equation}
where the parameter $\lambda$ is a weighting factor that controls the relative importance of trust versus user influence. 
As the trust $u_{ij}$ between agent $a_i$ and $a_j$ increases, the corresponding trust-based weight term $\frac{\lambda u_{ij}}{\sum_{k\in \mathcal{J}_{i,t}}u_{ik}}$ also increases. 
Similarly,  agents with more followers contribute more to the influence weight through the second term.  Clearly, $\omega_{ij}\geq 0$, and $\sum_{j=1}^{N}\omega_{ij}=1$. 

\textbf{Message Function}  $\mathcal{F}_{message}$ specifies the  message $m_{i,t+1}$ that agent $a_i$ broadcasts. 
Typically, this function assumes the message directly reflects the attitude $o_{i,t}$, and it serves as an input for attitude updates of its followers in the next timestep,
\begin{equation}
  \mathcal{F}_{message}: \, m_{i,t+1} = o_{i,t}.
\end{equation}

\textbf{Multi-Factor Relationships Predictor}\ 
To capture the evolution of relationships, we propose a Multi-Factor Relationships Predictor (MFRP) $\mathcal{F}_{predict}$, which simulates how agent \( a_i \) follows agent \( a_j \) from the candidate followees $\mathcal{N}_i$ and unfollows agent \( a_k \) from its following list \( \mathcal{G}_i^{ing} \), as depicted in Fig. \ref{fig:framework} (3.2):
\begin{equation}
[a_j, a_k] = \mathcal{F}_{predict}(a_i, \, \mathcal{N}_i,\,  \mathcal{G}_{i}^{ing}),
\end{equation}
where $\mathcal{N}_i$ is composed of followees-of-followees and high-follower non-neighbor agents with similar stances.

Social relationships are influenced by a variety of complex factors,  contributing to the formation and dissolution of ties. 
Concretely, users prefer connecting with like-minded peers while avoiding opposing views \cite{nickerson1998confirmation, sears1967selective}, represented by $|o_{i,t}-m_{j,t}|$. 
Moreover, social networks are inherently asymmetric due to the users influence, often proportional to their followers, denoted as $|\mathcal{G}_j^{er}|$.
Concurrently, one-way link $e_{ji}$ and trust $u_{ij}$ introduce fine-grained interactions into social relationships, reflecting the varying directions and strengths of social relationships. 
Based on these considerations, the predictor $\mathcal{F}_{predict}$ 
integrates the multi-dimensional factors to compute the metric $S_{ij}$, thereby capturing the dynamic nature of social relationships. As the follow and unfollow behaviors involve different target agents, they are treated as independent processes. Consequently, the metric $S_{ij}$ is formally defined in two distinct forms, $S_{ij}^{+}$ and $S_{ij}^{-}$:
\begin{equation}
\small
S_{ij}^{+} =  \frac{2-|o_{i,t} - m_{j,t}|}{2} + \frac{|\mathcal{G}_j^{er}|}{\max_{k \in \mathcal{N}_i} |\mathcal{G}_k^{er}|} + e_{ji},
\end{equation}
\begin{equation}
\small
S_{ij}^{-} = \underbrace{\frac{2 - |o_{i,t} - m_{j,t}|}{2}}_{\text{stance}} + \underbrace{\frac{|\mathcal{G}_j^{er}|}{\max_{k \in \mathcal{G}_i^{ing}}|\mathcal{G}_k^{er}|}}_{\text{influence}} + \underbrace{e_{ji}}_{\text{one-way}} + \underbrace{u_{ij}}_{\text{trust}},
\end{equation}
where higher values of the four terms indicate stronger interactions between agents, while lower values correspond to weaker connections. 
Moreover, unlike unfollow behavior, when agent $a_i$ establishes a new follow relationship with agent $a_j$, there is no pre-existing trust between them. 
A corresponding trust $u_{ij}$ should be formed through an intermediary agent $a_k$. 
Combining all potential trust propagation paths, the trust relationships $u_{ij}$ between agents can be expressed as:
\begin{equation}
\small
u_{ij}=\frac{\sum_{k \in \mathcal{G}_i^{ing}}\bar{u}_{ij}^k}{\sum_{k \in \mathcal{G}_i^{ing}}e_{ik}e_{kj}}, \, \overline{u}_{ij}^{k}=\frac{u_{ik}u_{kj}}{1+(1-u_{ik})(1-u_{kj})}.
\end{equation}

During the simulations, ordinary agents periodically perform the predictor $\mathcal{F}_{predict}$  from $T_{start}$ every $T_{interval}$ timesteps, with probabilities $p_{follow}$ and $p_{unfollow}$ for following and unfollowing agents, respectively.
When following, agent $a_i$ selects the agent $a_j$ from the candidate followees $\mathcal{N}_i$ with the highest \( S_{ij}^{+} \) as a new followee. Meanwhile, the agent $a_k$from the following list \( \mathcal{G}_i^{ing} \) with the lowest \( S_{ij}^{-} \) is selected for unfollowing,
\begin{equation}
    \begin{aligned}
    j = max_{a_j \in \mathcal{N}_i} S_{ij}^{+}, \quad
    k = min_{a_j \in \mathcal{G}_i^{ing}} S_{ij}^{-}.
  \end{aligned}
  \end{equation}

In this way, the dynamic-relationship-oriented ABM models the heterogeneous interaction among agents, closely mimicking real-world social behaviors where not all relationships are equal in influence or strength. 
By extending traditional ABMs with the multi-factor relationships predictor, ordinary agents capture the formation and dissolution of social relationships driven by multi-dimensional factors, thus improving the fidelity of attitude propagation and decision-making dynamics in simulations.

\section{Experiments}

\subsection{Experimental Settings}

\textbf{Configurations:} We use the \textit{GPT-4o-mini} to construct experiments, with \textit{text-embedding-3-large} used to obtain the embedding of tweet content. Simulation parameters are specified as follows, $\beta=-0.05$, $T_{start} = 1$, and $T_{interval} = 3$, which are adopted from related work \cite{DBLP:journals/corr/abs-2411-11581, DBLP:journals/isci/YangWDCXH24}. To reduce resource usage, most experiments involve 1,000 agents with 20 core agents across 12 simulation timesteps, while large-scale collective behaviors analysis  comprises 100,000 agents and 2,000 core agents. 
All experiments are executed on a server with 64 Intel(R) Xeon(R) Platinum 8352V CPU @ 2.10GHz and 240GB RAM.

\textbf{Datasets:}
To validate the effectiveness of $DynamiX$ in simulation dynamics, we select five famous events from Wikipedia: the Xinjiang Cotton\footnote{\url{en.wikipedia.org/wiki/Xinjiang_cotton_industry}}, Moon Landing Conspiracy\footnote{\url{en.wikipedia.org/wiki/Moon_landing_conspiracy_theories}},  Trump-Russia Investigation\footnote{\url{en.wikipedia.org/wiki/Russian_interference_in_the_2016_United_States_elections}}, Taiwan's stance on Military Parade of China's Victory Day\footnote{\url{en.wikipedia.org/wiki/2025_China_Victory_Day_Parade}}, and Euthanasia\footnote{\url{en.wikipedia.org/wiki/Euthanasia}}. We collect relevant tweets to build datasets, which include agent personas, IDs, follower counts, followee counts, tweet content, creation timestamps, and corresponding attitudes scored by \textit{GPT-4o-mini}. Five datasets are collected from the various domains and contextual scenarios, characterized by a long temporal span and a large volume of tweets, with details provided in Table \ref{datasets}. The Taiwan's stance on the Military Parade dataset, collected after the July 2024 knowledge cutoff of \textit{GPT-4o-mini}, aims to avoid data leakage issues. To ensure a systematic and effective evaluation of $DynamiX$, we conduct ablation and analysis experiments  on the Xinjiang Cotton, while performing comparisons with state-of-the-art methods and follower growth prediction experiments on the other datasets. 
Additionally, we use the Congress dataset \cite{FINK2023129083}  to evaluate the performance of link prediction.

\begin{table}[t]
  \centering
  \scriptsize
  \caption{Statistics of our datasets}
    \begin{tabular}{l|cccc}
    \toprule
    \multicolumn{1}{c|}{\textbf{Dataset}} & \textbf{Domain} & \textbf{\#Tweet} & \textbf{Start time} & \textbf{End time} \\
    \midrule
    Xinjiang Cotton  & Society & 14232 & Mar 15, 2020 & Sep 15, 2020\\
    Moon Landing & Technology & 6721 & Nov 01, 2022 & Nov 01, 2024 \\
    Trump-Russia  & Politics & 10335 & May 10, 2017 & Nov 10, 2017 \\
    Taiwan-Parade & Military & 20746 & Aug 14, 2025 &Sep 08, 2025\\
    Euthanasia & Ethics & 9663 & Jun  23, 2025 & Jul  25, 2025\\
    \bottomrule
    \end{tabular}%
  \label{datasets}%
\end{table}%


\textbf{Metrics:} To assess the accuracy of $DynamiX$ in simulating real-world attitude dynamics, we adopt a set of  widely accepted metrics \cite{DBLP:conf/acl/MouWH24,liu2025rumorsphereframeworkmillionscaleagentbased} to quantify the differences between the simulated and real-world public attitudes. 
Specifically, we measure \textit{Numerical Distribution} by evaluating $\Delta$Bias, which represents the mean deviations of the simulated public attitudes from the real world across different timesteps, and $\Delta$Div that indicates stability through the variance of the  deviations. 
We also use Dynamic Time Warping (DTW) \cite{muller2007dynamic} and Frechet distance \cite{eiter1994computing} to measure the \textit{Trend Shape} similarity between the simulated and real public attitudes. 
Additionally, to evaluate the simulation accuracy in modeling social relationships, we use F1, Precision, and Recall as metrics.

\begin{table*}[t]
  \centering
  \caption{Results of macro alignment evaluation. \colorbox{second_best}{best} and \colorbox{best}{second best} results are highlighted. The $\downarrow$ symbol indicates that smaller values correspond to a closer match with the real-world results.}

    \begin{tabular}{c|cccc|cccc|cccc}
      \toprule
      \multicolumn{1}{c|}{\multirow{2}[2]{*}{\textbf{Method}}} & \multicolumn{4}{c|}{\textbf{Moon Landing Conspiracy}} & \multicolumn{4}{c|}{\textbf{Trump-Russia Investigation}} & \multicolumn{4}{c}{\textbf{Taiwan-China Victory Parade}} \\
            & $\Delta$Bias$\downarrow$ & $\Delta$Div$\downarrow$ & DTW$\downarrow$ & Frechet$\downarrow$ & $\Delta$Bias$\downarrow$ & $\Delta$Div$\downarrow$ & DTW$\downarrow$ & Frechet$\downarrow$ & $\Delta$Bias$\downarrow$ & $\Delta$Div$\downarrow$ & DTW$\downarrow$ & Frechet$\downarrow$ \\
      \midrule
      BC\cite{DBLP:journals/advcs/DeffuantNAW00} & \cellcolor[rgb]{ .867,  .922,  .969}0.1211 & 0.1641 & 0.5803 & 0.3542 & 0.1162 & 0.0942 & 0.4514 & 0.2444 & \cellcolor[rgb]{ .867,  .922,  .969}0.0399 & 0.0585 & \cellcolor[rgb]{ .867,  .922,  .969}0.1867 & 0.1481 \\
      HK\cite{DBLP:journals/jasss/HegselmannK02} & 0.1357 & 0.1535 & \cellcolor[rgb]{ .867,  .922,  .969}0.3799 & \cellcolor[rgb]{ .867,  .922,  .969}0.2186 & 0.1760 & 0.1911 & 0.5349 & 0.2350 & 0.1022 & 0.0844 & 0.4198 & 0.2522 \\
      RA\cite{DBLP:journals/jasss/DeffuantAWF02} & 0.1588 & 0.1015 & 0.5540 & 0.2390 & 0.1383 & 0.0852 & 0.5447 & 0.2738 & 0.0431 & \cellcolor[rgb]{ .867,  .922,  .969}0.0532 & 0.2139 &0.1481 \\
      SJ\cite{DBLP:journals/cmot/JagerA05} & 0.2927 & 0.1138 & 0.9603 & 0.3909 & 0.2024 & 0.2252 & 0.5259 & 0.2648 & 0.0731 & 0.0564 & 0.2706 & \cellcolor[rgb]{ .867,  .922,  .969}0.1374 \\
      LR\cite{lorenz2021individual} & 0.2494 & 0.1736 & 0.9647 & 0.4531 & 0.3457 & 0.1576 & 1.2674 & 0.5932 & 0.1092 & 0.0855 & 0.3132 & 0.1576 \\
      \midrule
      FPS\cite{DBLP:conf/ijcai/LiuCZG0024} & 0.3188 & 0.1045 & 1.1055 & 0.3554 & 0.4273 & 0.1584 & 1.4522 & 0.5089 & 0.1771 & 0.1270 & 0.4949 & 0.2215 \\
      SOD\cite{DBLP:conf/naacl/ChuangGHSHYSHR24} & 0.1723 & \cellcolor[rgb]{ .867,  .922,  .969}0.0891 & 0.5974 & 0.2760 & \cellcolor[rgb]{ .867,  .922,  .969}0.0879 & \cellcolor[rgb]{ .867,  .922,  .969}0.0787 & \cellcolor[rgb]{ .867,  .922,  .969}0.3338 & \cellcolor[rgb]{ .867,  .922,  .969}0.2098 & 0.1845 & 0.1207 & 0.5743 & 0.2881 \\
      HiSim\cite{DBLP:conf/acl/MouWH24} & 0.1760 & 0.2183 & 0.6812 & 0.4203 & 0.1954 & 0.2113 & 0.7805 & 0.5139 & 0.1028 & 0.0743 & 0.3218 & 0.1666 \\
      \midrule
      Ours  & \cellcolor[rgb]{.988,  .894,  .839}0.0612 & \cellcolor[rgb]{.988,  .894,  .839}0.0605 & \cellcolor[rgb]{.988,  .894,  .839}0.1173 & \cellcolor[rgb]{.988,  .894,  .839}0.0586 & \cellcolor[rgb]{.988,  .894,  .839}0.0662 & \cellcolor[rgb]{.988,  .894,  .839}0.0657 & \cellcolor[rgb]{.988,  .894,  .839}0.2080 & \cellcolor[rgb]{.988,  .894,  .839}0.1330 & \cellcolor[rgb]{.988,  .894,  .839}0.0314 & \cellcolor[rgb]{.988,  .894,  .839}0.0450 & \cellcolor[rgb]{.988,  .894,  .839}0.1453 & \cellcolor[rgb]{.988,  .894,  .839}0.1229 \\
      \bottomrule
      \end{tabular}%
  \label{macro}%
\end{table*}%

\textbf{Baselines:} We verify the effectiveness of our method using two types of baselines.

\begin{itemize}[leftmargin=0.5em]
\item  \textbf{Agent-Based Models} define how agents adjust their attitudes based on the messages from their neighbors through predefined rules. We select five classic ABMs, including Bounded Confidence (BC) \cite{DBLP:journals/advcs/DeffuantNAW00}, Bounded Confidence Model-Multiple (HK) \cite{DBLP:journals/jasss/HegselmannK02}, Relative Agreement (RA) \cite{DBLP:journals/jasss/DeffuantAWF02}, Social Judgement (SJ)  \cite{DBLP:journals/cmot/JagerA05}, and LorenZ (LR) \cite{lorenz2021individual}. The  differences among them lie in the perspectives considered by the selection, update, and message functions, while the simulation process remains consistent. 

\item \textbf{LLM-based social network simulators} leverage the strong capabilities of LLMs in emulating complex individual behaviors and social dynamics. 
FPS \cite{DBLP:conf/ijcai/LiuCZG0024} captures semantic nuances and models the spread of fake news, replicating opinion dynamics and analyzing intervention strategies. SOD \cite{DBLP:conf/naacl/ChuangGHSHYSHR24} emulates belief formation and confirmation bias, effectively mirroring human cognition and collective consensus.  HiSim \cite{DBLP:conf/acl/MouWH24} integrates LLM-based agents with traditional ABMs to scale up simulation populations, offering insights into individual behaviors and large-scale social dynamics. 
\end{itemize}


\subsection{Comparisons with State-Of-The-Art Methods}
\label{Macro Alignment Evaluation}

\textbf{Finding 1: Compared to static simulators, $DynamiX$ effectively captures the dynamic evolution of public attitudes, exhibiting advantageous accuracy, stability, and adaptability across different events, including future, yet-to-occur events.} 

To validate the alignment with real-world attitudes, we systematically conduct a Macro Alignment Evaluation to compare the simulated results of $DynamiX$ with other static baselines.
As summarized in Table \ref{macro}, $DynamiX$ consistently maintains an optimal performance on all metrics across three events with different propagation patterns, underscoring its accuracy, robustness, and generalizability in cross-domain event alignment. 
In contrast, existing studies, which typically rely on fixed agent roles and static social networks, fail to accurately model attitude dynamics, further emphasizing the importance of simulating the dynamic nature of social networks.

Specifically, in terms of \textit{Numerical Distribution}, $DynamiX$ achieves a reduction in the mean values of the second-best model by 0.0300 $\Delta$Bias and 0.0166 $\Delta$Div, thereby substantiating its advantage in marked stability and accurate reflection of public attitude dynamics over long-time simulations. 
Regarding \textit{Trend Shape} alignment, $DynamiX$ exhibits notable improvements in both DTW and Frechet metrics, realizing respective average decreases of 0.2322 and 0.0736 relative to the second-best model. This affirms its effectiveness in accurately capturing temporal, nonlinear characteristics inherent in attitude evolution.
Moreover, the results from the Taiwan-China Victory Day Parade event maintain accurate alignment with real-world data, demonstrating its ability to infer the dynamic evolution of attitudes towards future, yet-to-occur events, due to the dataset collected after the July 2024 knowledge cutoff of \textit{GPT-4o-mini}.

\begin{figure}[t]
  \setlength{\abovecaptionskip}{0.1cm}
  \includegraphics[width=0.9\linewidth]{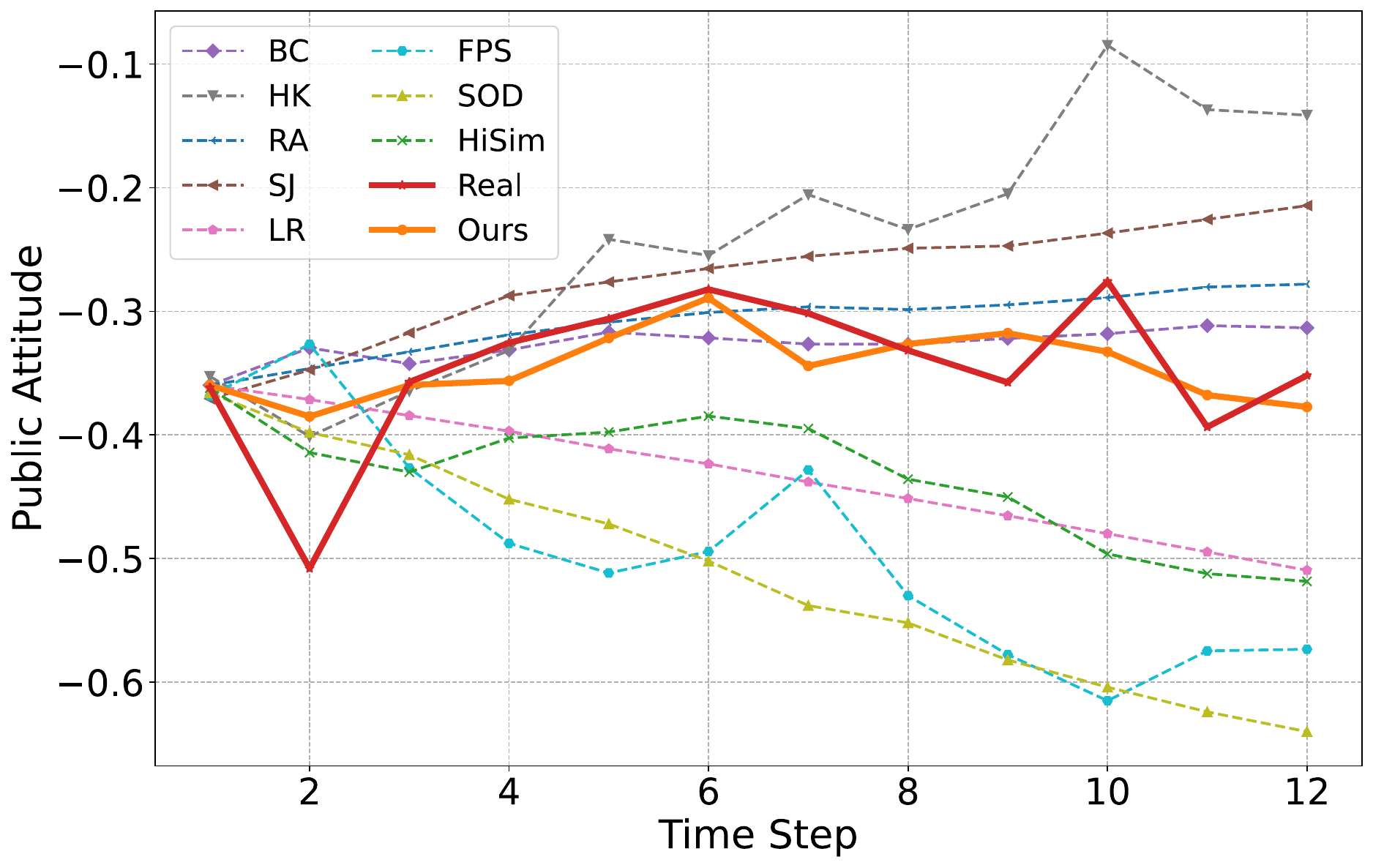}
\caption{Visualization of the different evolution results on Taiwan-China Victory Day Parade event. $DynamiX$ demonstrates a high degree of alignment and consistency with the real-world dynamics.}
\label{visualization}
\end{figure}

For an intuitive comparison of the differences between the simulated and real-world attitude dynamics, we present the visualization results of the Taiwan-China Victory Day Parade in Fig. \ref{visualization}. 
Consistent with the preceding analysis, $DynamiX$ exhibits a remarkable alignment with real-world dynamics. 
In contrast, traditional ABMs, which rely solely on initial attitudes and defined interaction rules, encounter significant challenges in replicating propagation patterns.
This is primarily due to their inability to model the heterogeneous cognition of users and to incorporate event descriptions within the simulation process.
Meanwhile, LLM-based baselines face difficulties in integrating novel viewpoints due to the reliance on fixed core agents, leading to the entrenchment of pre-existing stances.
As a result, their simulated social dynamics tend to converge rapidly around core agents with extreme attitudes, failing to capture the large-scale evolutionary patterns observed in the real world. 
Through modeling the evolving roles of core agents and dynamic social relationships, $DynamiX$ achieves superior performance and demonstrates strong adaptability and stability across different timesteps, confirming its effectiveness in simulating and analyzing the dynamics of large-scale social simulations.

\begin{figure}[t]
  \centering
\setlength{\abovecaptionskip}{0.1cm}
\includegraphics[width=1\linewidth, height = 0.6\linewidth]{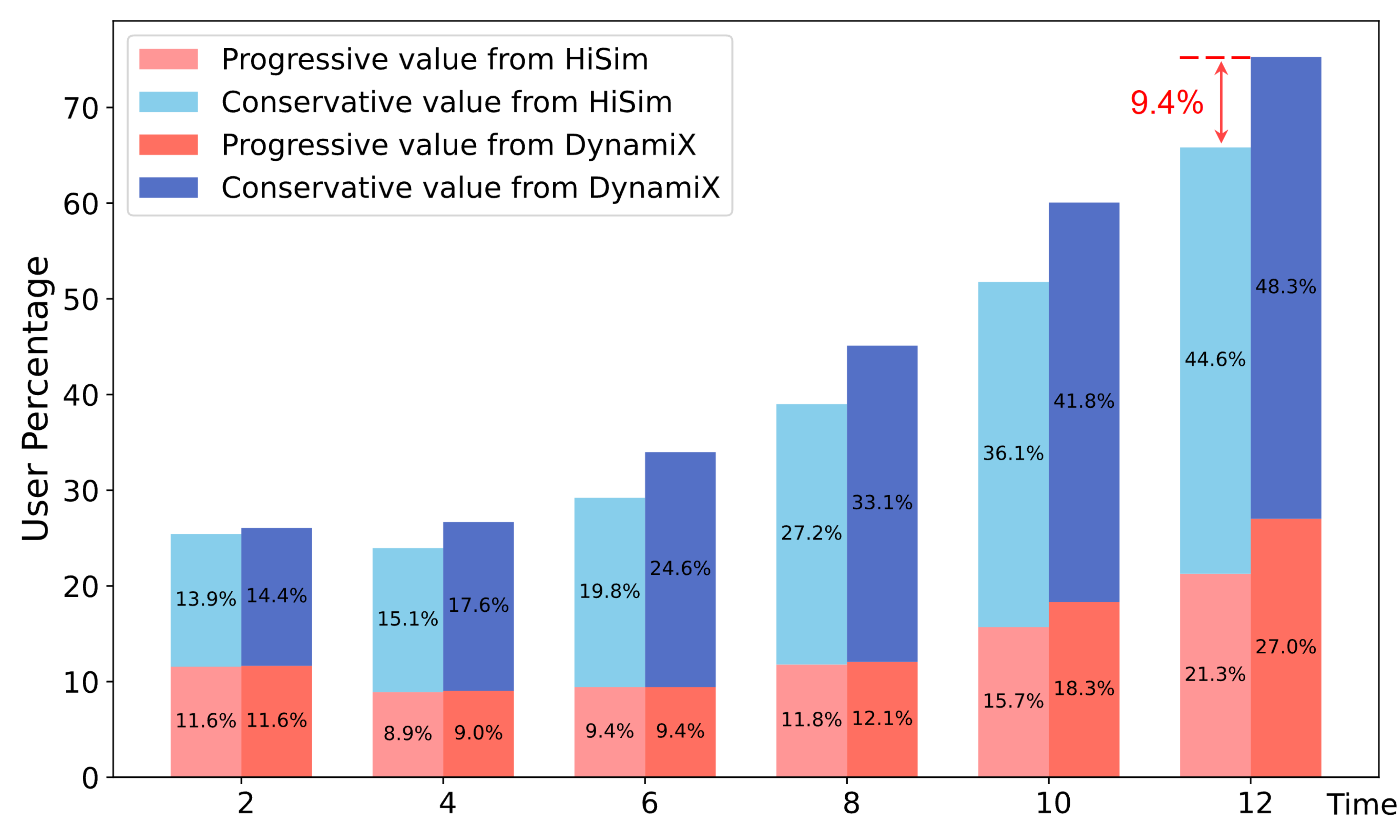}
  \caption{An illustration of extreme attitudes across timesteps, showing $DynamiX$ significantly accelerates polarization compared to HiSim. }
  \label{deco}
\end{figure}

\textbf{Finding 2: Compared to static simulators, $DynamiX$ accelerates attitude polarization significantly, better reflects real-world propagation patterns and collective behaviors.} 

To assess the influence of dynamic social networks upon large-scale collective behaviors, we simulate the attitude evolution of 100,000 agents concerning Euthanasia event.  
We then execute a comparative analysis between $DynamiX$
and static social networks simulator \emph{i.e.,} HiSim \cite{DBLP:conf/acl/MouWH24}.
We define the Progressive, Conservative, and Draw metrics as the percentages of users with attitudes greater than 0.4, less than -0.4, and within [-0.4, 0.4], respectively, along with the Variance, which quantifies the overall variance in public attitudes.

As depicted in Fig. \ref{deco}, the incorporation of the switching agent roles and dynamic social relationships significantly accelerates the polarization process, with a notable increase in the proportion of agents exhibiting extreme attitudes. 
Concretely, the percentage of extreme attitude rises from 34.0\% to 75.3\% in $DynamiX$, an increase of 41.3\% between timestep 6 and 12, surpassing the 36.7\% polarization increase observed in HiSim by 4.6\%. 
Meanwhile, $DynamiX$ exhibits 9.4\% more polarized attitudes than HiSim, highlighting the pivotal roles that dynamic social networks play in driving collective attitude polarization, consistent with existing sociological studies \cite{DBLP:journals/jocss/SasaharaCPCFM21, DBLP:journals/pnas/SantosLL21}.

\begin{figure}[t]
  \centering
  \setlength{\abovecaptionskip}{0.1cm}
 \includegraphics[width=1\linewidth, height = 0.6\linewidth]{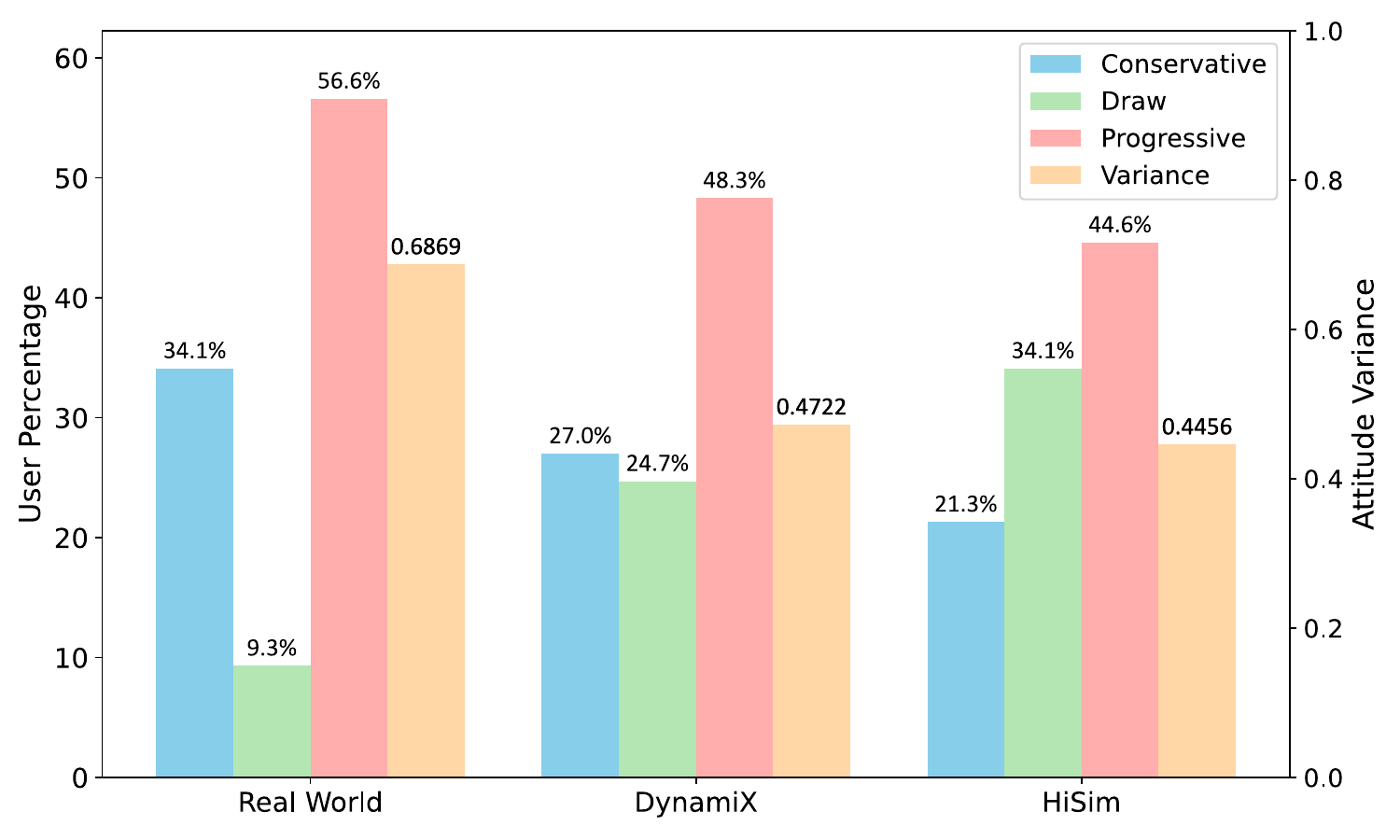}
  \caption{Comparisons of attitude polarization on Euthanasia across real world, $DynamiX$, and HiSim.}
  \label{Euthanasia_label}
\end{figure}

As an ongoing topic with relatively stable public opinions, Euthanasia serves as an ideal subject for comparing attitude polarization results. 
Therefore, we compare the simulated results with the real world at the final timestep to validate the robustness and validity of $DynamiX$ in replicating collective behaviors.
As shown in the Fig. \ref{Euthanasia_label},
$DynamiX$ yields polarization outcomes and attitudes variance that more closely align with the real world, with more consistent performance of 3.7\% in Progressive, 5.7\% in Conservative, 9.4\% in Draw, and 0.0266 in Variance compared to HiSim.
The result demonstrates $DynamiX$ achieves improved alignment with real-world propagation dynamics, emphasizing the effectiveness, robustness, and realism of dynamic social relationships in  accelerating attitude polarization.

\begin{figure*}[t]
	\centering

	\begin{minipage}[t]{0.48\linewidth}
		\centering
		\includegraphics[width=\linewidth]{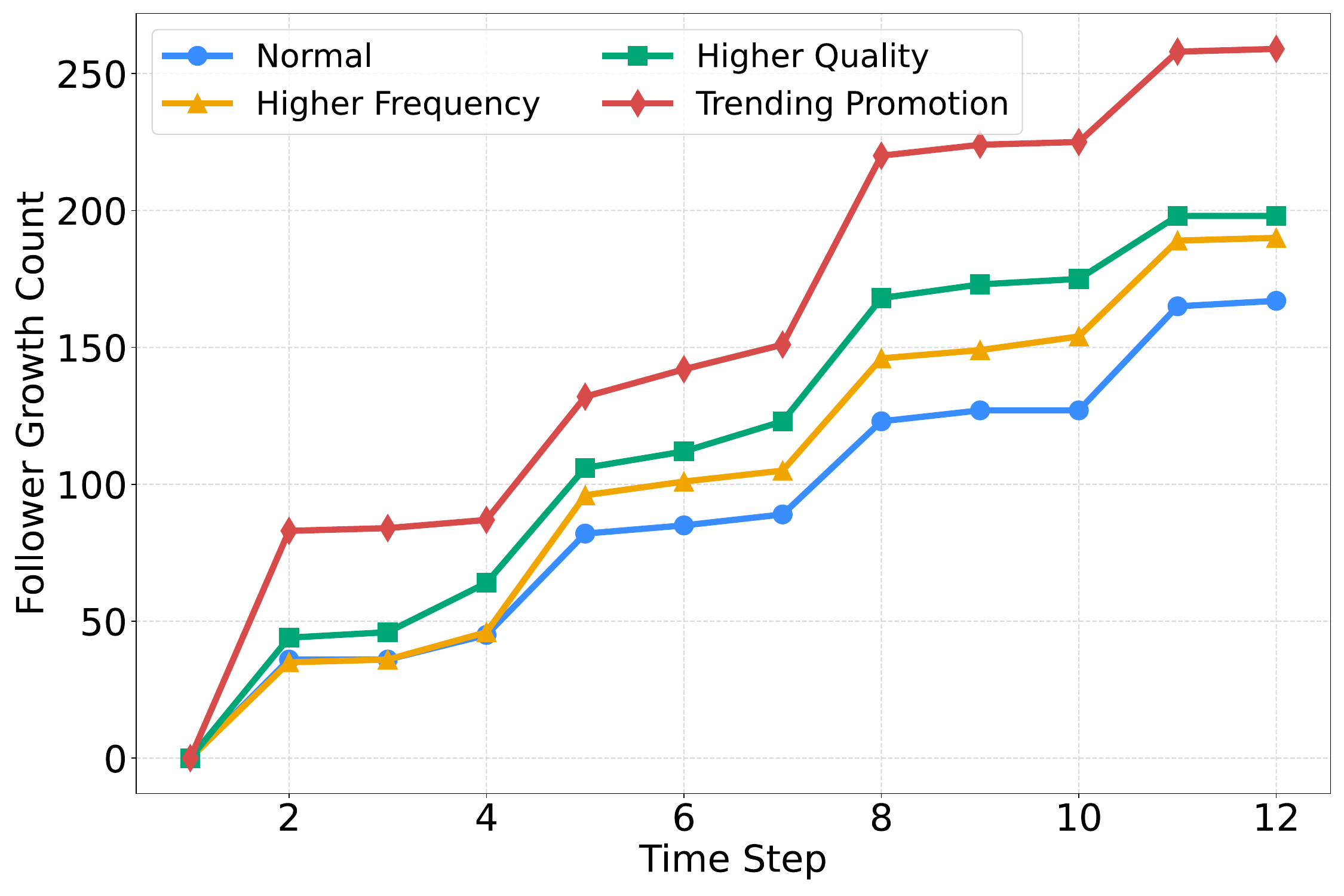}
    \small
		(a) Follower growth of high-influence users
	\end{minipage}
	\hfill
	\begin{minipage}[t]{0.48\linewidth}
		\centering
		\includegraphics[width=\linewidth]{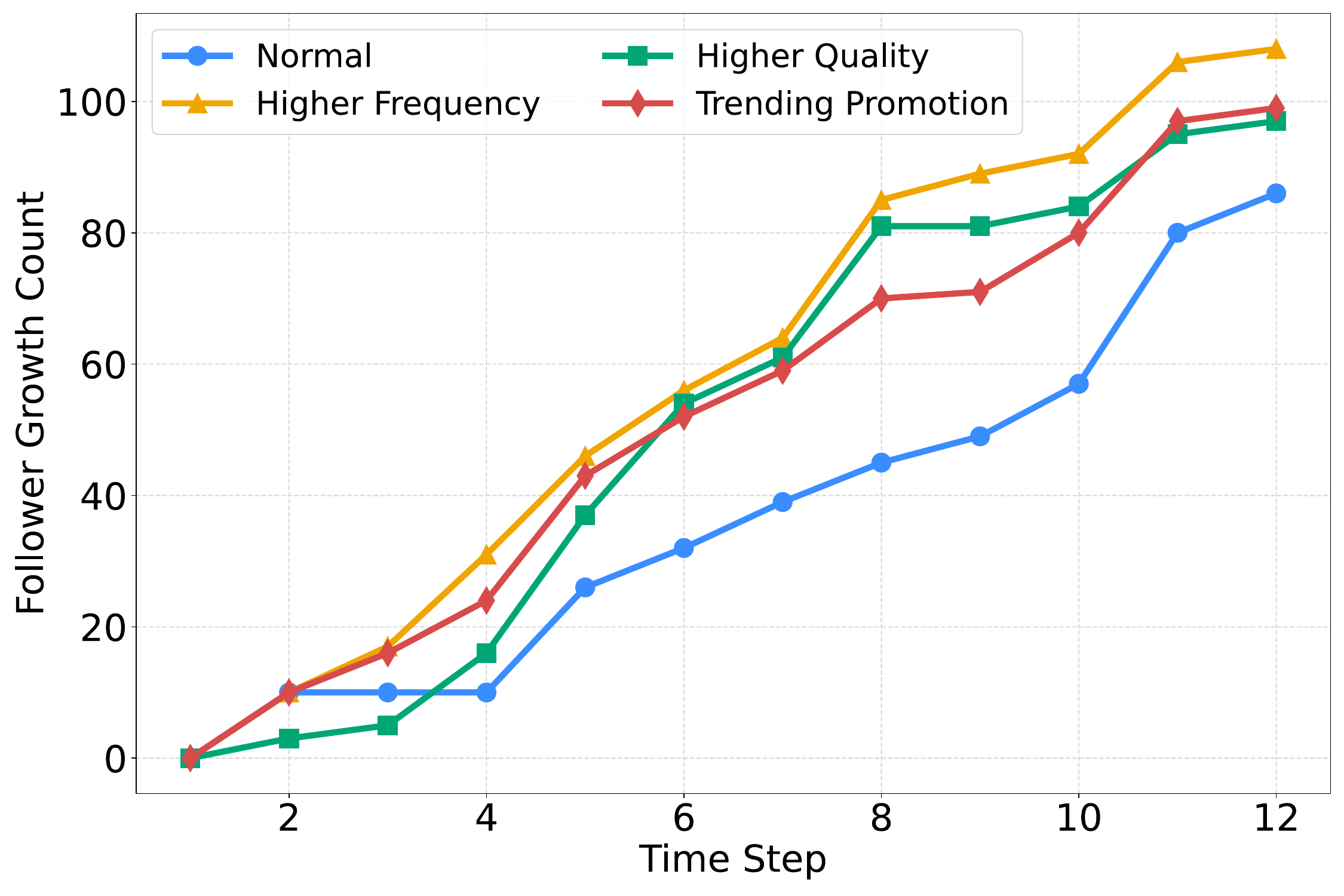}
    \small
		(b) Follower growth of low-influence users
	\end{minipage}
	
	\caption{Comparisons of follower growth prediction under different measures across user categories.}
	\label{fig:high_low_influence}
\end{figure*}

\subsection{Follower Growth Prediction}
\label{Follower Growth Prediction}

\textbf{Finding 3: $DynamiX$ facilitates accurate prediction of follower growth during event propagation, providing empirical support for the cultivation of opinion leaders.}  

Through modeling the formation and dissolution of  social relationships, $DynamiX$ makes it possible to predict follower growth, thereby opening up a new perspective to study social simulation. In this context, higher tweet frequency, higher tweet quality, and trending promotion are widely recognized as key measures influencing follower growth in real-world social networks. Therefore, we conduct an experiment to investigate the effects of these three measures on follower growth dynamics,  including high-influence users and low-influence users. These measures are implemented by adjusting the behaviors decision of core agents, as only core agents engage in interactive dialogues through tweets. 

\begin{itemize}[leftmargin=0.5em]
\item For higher quality, we design prompts that encourage agents to post higher-quality tweets with longer, more opinionated, and attention-grabbing content, simulating their impact on follower growth. The prompts are provided in the Appendix. 
\item To achieve a higher frequency, we increase the probability $\phi_{i,t}$ of an agent being selected as a core agent. As this probability increases, the frequency of tweet posting rises, which in turn affects follower growth.
\item To simulate trending promotion, we increase the probability of tweets appearing on personalized information streams $\boldsymbol{R}_j$ of the agent $a_j$. This way, the tweet is more likely to be seen by the broader audience, thus boosting follower growth.
\end{itemize}

As illustrated in Fig. \ref{fig:high_low_influence}, the comparisons of follower growth patterns under different measures reveal the following: 1) Follower growth for high-influence users is more easily achieved through trending promotion, whereas low-influence users predominantly benefit from continuously sharing high-quality content to enhance follower retention. 2) Higher frequency has a limited effect on follower growth for high-influence users. This is likely because they already have high exposure and relatively fixed follower circles. Simply increasing frequency does not significantly enhance effectiveness in gaining additional growth beyond their existing followers. Trending promotion, however, is an effective way to break through the boundaries of these follower circles and reach a new audience. 3) Higher quality consistently promotes significant follower growth across both user categories, indicating content quality is the core driver of the decision-making process on social platforms. High-quality content meeting inner needs can achieve steady follower growth. 4) Trending promotion demonstrates an insignificant effect on follower growth for low-influence users, because the audience remains skeptical about their influence, resulting in a slower growth. 

To the best of our knowledge, $DynamiX$ is the first attempt to predict follower growth within large-scale social network simulators. The above results demonstrate the ability of $DynamiX$ in forecasting follower growth,  highlighting its substantial commercial potential in cultivating opinion leaders.


%

\begin{figure}[t]
  \centering
  \setlength{\abovecaptionskip}{0.1cm}
 \includegraphics[width=1\linewidth]{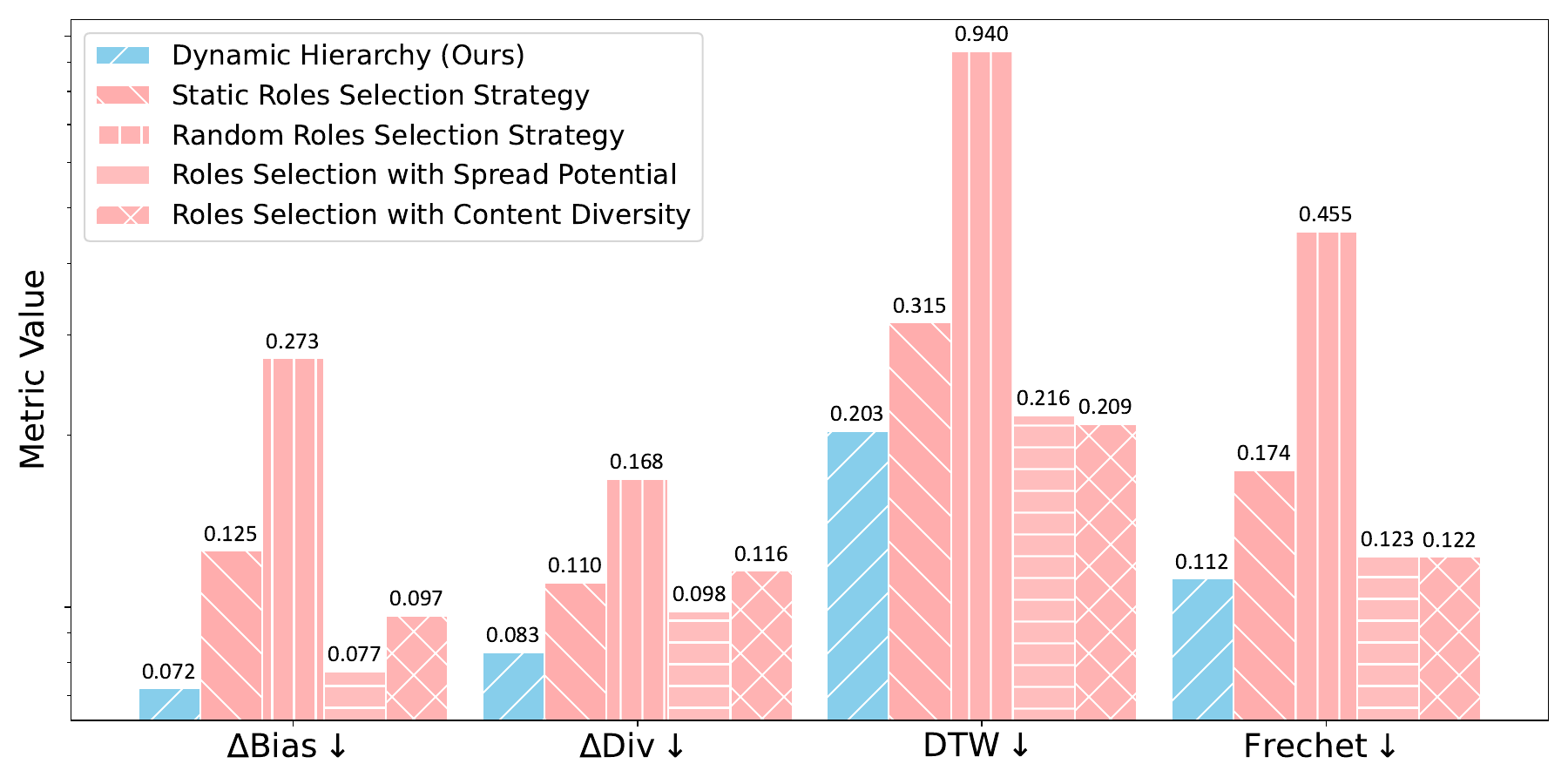}
  \caption{The contribution illustration of  Dynamic Hierarchy  module.}
  \label{ablation3}
\end{figure}

\subsection{Ablation Study}
\label{Ablation Study}

To evaluate the contributions of each component on attitude evolution, we conduct ablation experiments on the Xinjiang Cotton dataset.

\textbf{Agent Roles Assignment:} To evaluate the effectiveness of role-switching, we replace the dynamic hierarchy module with the selection strategies used in existing simulators, with the results shown in Fig. \ref{ablation3}. 
When replacing DH with fixed or random strategies, the performance in predicting attitude evolution degrades significantly, with differences of 0.053 and 0.201 in $\Delta$Bias, and 0.027 and 0.085 in $\Delta$Div, respectively.
This result confirms that the accuracy of existing simulators is greatly affected when the number of core agents is finite. 
The strategy that considers only spread influence or content diversity shows slight performance degradation, but still outperforms the fixed and random strategies. 
In contrast, our dynamic hierarchy, which accounts for both factors to select core agents, strikes a balance between simulation accuracy and scalability, successfully replicating real-world attitude dynamics, even with a finite number of core agents.

\textbf{Dynamic Social Relationships:} To assess the impact of dynamic social relationships on attitude dynamics, we conduct an ablation study by removing the Personalized Relationships Recommendation (PRR) from core agents module and the Multi-Factor Relationships Predictor (MFRP) from ordinary agents module. 
As shown in Table \ref{ablation2}, $DynamiX$ achieves superior performance across all metrics. 
Specifically, the removal of PRR results in a performance decline, with increases of 0.0320 in $\Delta$Bias and 0.0195 in $\Delta$Div, indicating that the dynamic social relationships of core agents play a significant role in attitude dynamics. The MFRP component yields similar findings. 
Moreover, the social relationships without either follow or unfollow cause performance degradation in both the PRR and MFRP components. 
Follow plays a more significant role than unfollow behavior during the event propagation process, which is consistent with its more frequent occurrence in real-world society. 
Excluding follow results in worse performance, with a difference of 0.0028 and 0.0119 in $\Delta$Bias compared to unfollow for PRR and MFRP, respectively.
Furthermore, the results without both components present the most substantial performance degradation,  particularly the increases of  0.1884 in DTW and 0.0859 in Frechet, emphasizing the critical significance of dynamic social relationships in reproducing the  social dynamics trend across different user types. 
Together, these results highlight the formation and dissolution of social relationships facilitate alignment with real-world attitude dynamics.

\begin{table}[t]
  \centering
  \caption{Component ablations of attitude dynamics. }
    \begin{tabular}{cc|cccc}
      \toprule
      \multicolumn{2}{c|}{Model} & $\Delta$Bias$\downarrow$ & $\Delta$Div$\downarrow$ & DTW$\downarrow$ & Frechet$\downarrow$ \\
      \midrule
      \multicolumn{2}{c|}{ Ours} & \textbf{0.0720} & \textbf{0.0834} & \textbf{0.2035} & \textbf{0.1122} \\
      \midrule
      \multicolumn{1}{c|}{\multirow{3}[2]{*}{PRR}} & w/o  All & 0.1040 & 0.1029 & 0.2715 & 0.1750 \\
      \multicolumn{1}{c|}{} & w/o Follow & 0.0922 & 0.0931 & 0.2655 & 0.1407 \\
      \multicolumn{1}{c|}{} & w/o Unfollow & 0.0894 & 0.0964 & 0.2732 & 0.1502 \\
      \midrule
      \multicolumn{1}{c|}{\multirow{3}[2]{*}{MFRP}} & w/o All & 0.1102 & 0.1211 & 0.3207 & 0.1660 \\
      \multicolumn{1}{c|}{} & w/o Follow & 0.0976 & 0.1144 & 0.3271 & 0.1570 \\
      \multicolumn{1}{c|}{} & w/o Unfollow & 0.0857 & 0.0875 & 0.2532 & 0.1570 \\
      \midrule
      \multicolumn{2}{c|}{ w/o PRR \& MFRP} & 0.1315 & 0.0962 & 0.3919 & 0.1981 \\
      \midrule
      \multicolumn{2}{c|}{w/o inequality in ABMs} & 0.1179 & 0.1093 & 0.3218 & 0.1853 \\
      \bottomrule
      \end{tabular}%
  \label{ablation2}%
\end{table}

\begin{table}[t]
  \centering
  \caption{Consumption Comparisons of Tokens and Time}
  \begin{tabular}{c|c|ccccc}
    \toprule
     & $Token_{out}$  & $T_{sele} (min)$ & $T_{core} (min)$ & $T_{ordi} (min)$  \\
  \midrule
   Ours  &  57779  & 0.0017  & 30.6149  & 0.2475    \\
    Static  &  32662  & 0.0000  & 22.3473 & 0.2471   \\
  \bottomrule
  \end{tabular}
  \label{Consumption}
\end{table}%

\textbf{Unequal Interactions in ABMs:} To evaluate the effectiveness of unequal interactions, we substitute our DRO-ABM with traditional ABMs, where selected neighbors are treated equally in terms of attitude updates. 
As shown in Table \ref{ablation2}, $DynamiX$ exhibits a notable performance deterioration across all evaluation metrics. 
This result confirms the emergence of unequal interactions and highlights the critical role of DRO-ABM in accurately modeling attitude evolution.


\begin{table}[t]
  \centering
  \fontsize{8}{7.9}\selectfont
  \caption{Parameter sensitivity analysis of $p_{follow}$ and $p_{unfollow}$}
  \vspace{-0.3em}
    \begin{tabular}{l|cccc}
    \toprule
    \multicolumn{1}{c|}{Model} & $\Delta$Bias$\downarrow$ & $\Delta$Div$\downarrow$ & DTW$\downarrow$ & Frechet$\downarrow$ \\
    \midrule
    $p_{follow}$ = 0.0 &  0.0976 & 0.1144 & 0.3271 & 0.1570  \\
    $p_{follow}$  = 0.05 & 0.0961 & 0.1050 & 0.2203 & 0.1148 \\
    \textbf{$p_{follow}$  = 0.1}  & \textbf{0.0720} & \textbf{0.0834} & \textbf{0.2035} & \textbf{0.1122} \\
    $p_{follow}$  = 0.2 & 0.0724 & 0.0892 & 0.2249 & 0.1204  \\
    $p_{follow}$  = 0.4 & 0.0754 & 0.0870 & 0.2140 & 0.1223  \\
    $p_{follow}$  = 0.6 & 0.0781 & 0.0888 & 0.2866 & 0.1526  \\
    $p_{follow}$  = 0.8 & 0.1357 & 0.1155 & 0.3597 & 0.2029  \\
    \midrule
    $p_{unfollow}$ = 0.0 & 0.0857 & 0.0875 & 0.2532 & 0.1570 \\
    \textbf{$p_{unfollow}$  = 0.05}  & \textbf{0.0720} & \textbf{0.0834} & \textbf{0.2035} & \textbf{0.1122} \\
    $p_{unfollow}$ = 0.1 & 0.0756 & 0.0905 & 0.2076 & 0.1127 \\
    $p_{unfollow}$ = 0.2 & 0.0845 & 0.1068 & 0.2122 & 0.1197 \\
    $p_{unfollow}$ = 0.4 & 0.1349 & 0.1161 & 0.2871 & 0.1637 \\
    \bottomrule
    \end{tabular}%
  \label{Parameter}%
\end{table}%


\begin{figure*}[ht]
  \centering
  \setlength{\abovecaptionskip}{-0.1cm}
  \includegraphics[width=0.95\linewidth]{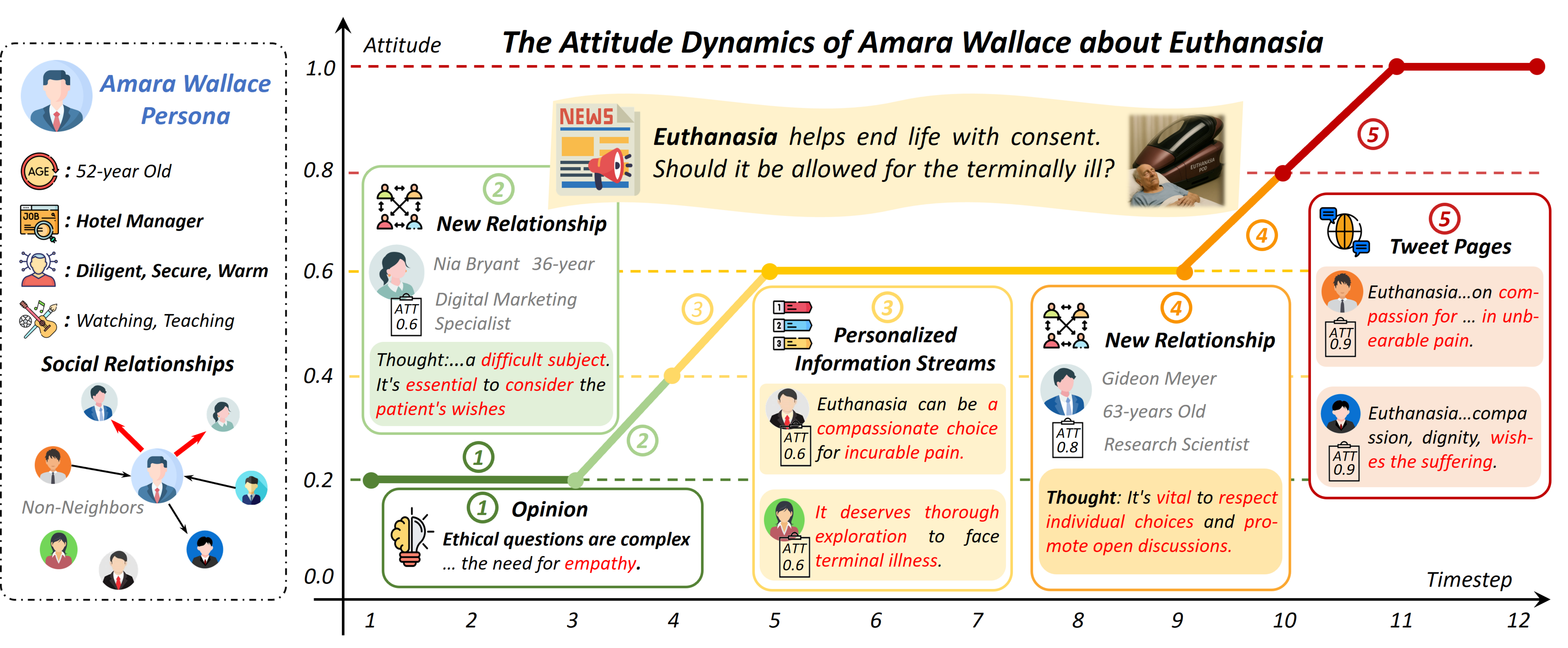}
  \caption{The evolution of individual attitudes towards Euthanasia within $DynamiX$. The stance of Amara Wallace shifts from neutral to progressive as he is continuously influenced by new relationships, personalized information streams, and tweet pages.}
  \label{micro}
\end{figure*}

\subsection{Sensitivity Analysis}
\label{Sensitivity Analysis}

\textbf{Consumption Analysis:} To compare the resource consumption of $DynamiX$ and  static simulator without the role-switching and dynamic social relationships, we track several key metrics during the simulation process: the output tokens, and the time spent on each module (in minutes).
The experimental configurations are consistent with those used in the macro alignment evaluation with 1,000 agents.
As shown in Table \ref{Consumption}, $DynamiX$ incurs only marginal increases in computational overhead, with  significant advantages over static simulator in predicting  attitude trends, replicating collective behaviors, and cultivating opinion leaders.
To be more precise, the token consumption of $DynamiX$ has increased by less than one time (0.77$\times$) compared to static simulator.  
This increase is mainly attributed to the additional decision-making processes regarding social relationships adjustments of the core agents, with their time consumption increasing by 0.36$\times$.
Furthermore, the time consumption in other modules for $DynamiX$ is similar to that of static networks. 
This indicates that the MFPR component for periodically adjusting the social relationships of ordinary agents incurs minimal computational cost.
And the dynamic hierarchy module addresses the limitations of existing selection strategies in balancing simulation performance and finite populations, requiring only 0.0017 minutes of computational time. 

\textbf{Parameter Analysis of $p_{follow}$ and $p_{unfollow}$:} The Multi-Factor Relationships Predictor governs the formation and dissolution of social relationships for ordinary agents. To systematically determine the optimal parameter configurations, we conduct an analysis experiment to optimize the parameters $p_{follow}$ and $p_{unfollow}$. As shown in Table \ref{Parameter}, the results exhibit a typical inverted U-shaped trend with respect to both parameters. When $p_{follow}$ = 0, ordinary agents cannot follow non-neighbors agents, resulting in a performance decrease of 0.0256 in $\Delta$Bias and 0.0310 in $\Delta$Div, thereby underscoring the importance of new follow relationships in enhancing the effectiveness of aligning with real-world attitude dynamics. When $p_{follow}$ falls within the range of 0.05 to 0.40, $DynamiX$ remains stable, and reaches its peak at $p_{follow}$ = 0.10. However, further increasing $p_{follow}$ beyond this range introduces excessive spurious links into the networks, resulting in poorer capability of modeling social relationships formation. Similarly, $DynamiX$ achieves favorable performance when $p_{unfollow}$ is between 0.05 and 0.20, peaking around $p_{unfollow}$ = 0.05. Consequently, experiment configurations adopt  $p_{follow}$ = 0.10 and $p_{unfollow}$ = 0.05 based upon the above empirical considerations.

\begin{table}[t]
  \centering
  \caption{Performance comparison of  relationships prediction models.}
  \fontsize{7.5}{8.5}\selectfont
  \begin{tabular}{l|c|ccc}
  \toprule
  Task Type & \multicolumn{1}{c|}{Value} & F1 $\uparrow$   & Precision $\uparrow$ & Recall $\uparrow$ \\
  \midrule
  \multirow{5}{*}{\shortstack{Follow\\Prediction}} 
  & Ours  & \textbf{0.2100} & \textbf{0.1682} & \textbf{0.2796} \\
  & CN \cite{DBLP:conf/cikm/Liben-NowellK03}   & 0.1345 & 0.1075 & 0.1796 \\
  & AA  \cite{DBLP:journals/socnet/AdamicA03}   & 0.1373 & 0.1098 & 0.1834 \\
  & Katz \cite{DBLP:journals/jasis/Liben-NowellK07} & 0.1532 & 0.1225 & 0.2046 \\
  & LPOD \cite{pan2016predicting} & 0.1886 & 0.1507 & 0.2518 \\
  \midrule
  \multirow{5}{*}{\shortstack{Unfollow\\Prediction}} 
  & Ours  & \textbf{0.5726} & \textbf{0.4593} & \textbf{0.7601} \\
  & CN \cite{DBLP:conf/cikm/Liben-NowellK03}   & 0.2537 & 0.2035 & 0.3368 \\
  & AA \cite{DBLP:journals/socnet/AdamicA03}   & 0.4263 & 0.3420 & 0.5660 \\
  & Katz \cite{DBLP:journals/jasis/Liben-NowellK07} & 0.5633 & 0.4518 & 0.7476 \\
  & LPOD \cite{pan2016predicting} & 0.5607 & 0.4496 & 0.7447 \\
  \bottomrule
  \end{tabular}
  \label{evaluation}
\end{table}

\textbf{Relationships Prediction Performance:} To evaluate the effectiveness of $DynamiX$ in predicting real-world social relationships, we randomly perturb 30\% of the edges in the Congress dataset. 
We then conduct the simulations of $DynamiX$ within 12 timesteps, while comparative methods perform relationships prediction tasks over the same timesteps. 
As shown in Table \ref{evaluation},  $DynamiX$ consistently outperforms all baseline methods across all metrics, surpassing the second-best model by 0.0214 and 0.0119 in F1 for predicting follow and unfollow behaviors, respectively. 
Traditional link prediction models focus mainly on structural information and are not designed for social network simulation.
They fail to account for the event-specific descriptions, and neglect multi-dimensional factors that drive the formation and dissolution of relationships, limiting their interpretability and generalizability in long-term, high-fidelity simulations. 
In contrast, $DynamiX$ incorporates a personalized relationships evolution engine for core agents, and a multi-factor relationships predictor for ordinary agents, 
addressing the above limitations to accurately model the dynamic mechanisms of  social relationships.
These results underscore the robustness and superior capability of $DynamiX$ in capturing the intricate mechanisms underlying the formation and dissolution of social relationships.


\textbf{Stability Analysis:} To assess the stability of $DynamiX$, we conduct variance measurements across multiple random seeds on the Xinjiang Cotton dataset. 
Specifically, we perform five independent runs for each method, and compute the mean and standard deviation of the metrics to evaluate their stability and robustness. 
As shown in Table \ref{stability}, $DynamiX$ consistently demonstrates superior average values with low variability across all metrics, highlighting its robustness and consistent performance across different random seeds.

\begin{table}[t]
  \centering
  \fontsize{7.5}{8.8}\selectfont
  \caption{Stability Analysis of DynamiX across multiple random seeds}
  \begin{tabular}{c|cccc} 
    \toprule
     & $\overline{\Delta Bias}\downarrow$  & $\overline{\Delta Div}\downarrow$ & $\overline{DTW}\downarrow$ & $\overline{Frechet}\downarrow$ \\
    \midrule
    BC& 0.224 (0.012)&0.152 (0.021)&0.761 (0.143)&0.380 (0.059) \\
    HK&0.451 (0.025)&0.258 (0.015)&1.794 (0.098)&0.765 (0.077) \\
    RA&0.278 (0.008)&0.223 (\textbf{0.005})&1.234 (0.023)&0.570 (0.013) \\
    SJ &0.175 (0.018)&0.135 (0.009)&0.626 (0.076)&0.328 (0.033) \\
    LR &0.482 (0.005)&0.287 (0.021)&1.945 (0.039)&0.831 (0.084) \\
    \midrule    
    FPS   & 0.327 (0.012)  & 0.136 (0.013)  &  0.803 (0.057)  &  0.309 (0.025)  \\    
    SOD   & 0.255 (0.023)  &  0.155 (0.012)  & 0.927 (0.109)  & 0.442 (0.038) \\     
    HiSim & 0.185 (0.016)  &  0.166 (0.026) &  0.797 (0.087)  &  0.393 (0.030)  \\     
    \midrule    
    Ours  & \textbf{0.073} (\textbf{0.003}) & \textbf{0.091} (0.007) & \textbf{0.209} (\textbf{0.006}) & \textbf{0.110} (\textbf{0.009}) \\ 
    \bottomrule 
  \end{tabular}
  \label{stability}
\end{table}

\subsection{Visualizations}

To reveal how $DynamiX$ promotes the evolution of individual attitudes, we present a case study in large-scale collective behaviors experiments, as shown in Fig. \ref{micro}. 
Amara Wallace initially holds a neutral attitude due to intrinsic personality traits \emph{e.g.,} warmth, responsibility. 
After establishing a new relationship with the person advocating the view that ``It's essential to consider the patient's wishes", he begins to adopt a moderately accepting stance towards Euthanasia. 
Subsequently, through exposure to homogeneous information via personalized information streams, his attitude further shifts towards a more extreme position. 
Ultimately, the tweets received from his neighbors reinforce his stance, leading him to advocate for ``end-of-life autonomy." 
In conclusion, sequential influences from new relationships, personalized information streams, and tweet pages drive the attitude transitions from moderate to extreme, revealing the intrinsic significance of dynamic social networks to micro mechanism in individual polarization and attitude dynamics.

\begin{figure}[t]
  \centering
  \setlength{\abovecaptionskip}{-0.1cm}
  {
  \includegraphics[width=0.9\linewidth]{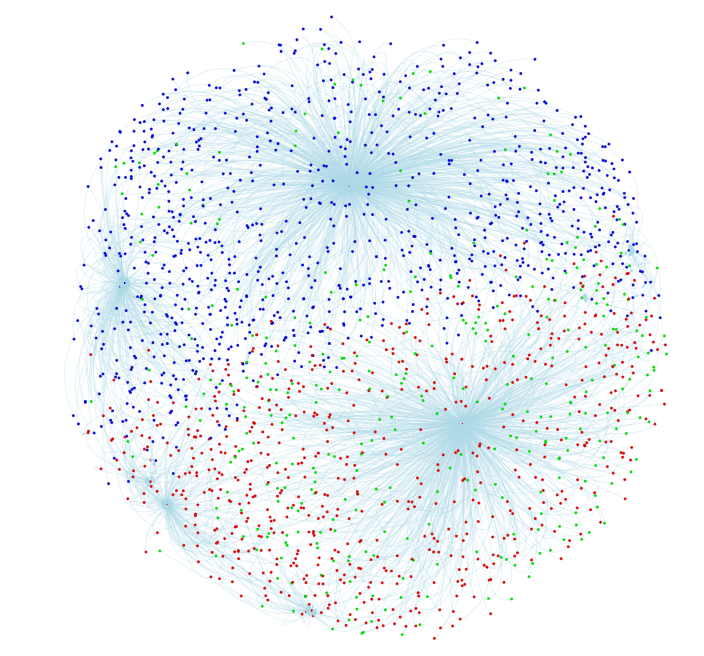}
  
  }
  \caption{New follow relationships exhibit clear clustering patterns, where nodes represent agents and edges represent new follow relationships. Colors indicate attitudes, with green for neutral, blue for conservative, and red for progressive.
  }
  \label{network}
\end{figure}

Finally, we visualize the results of new follow relationships during dynamic network evolution, which exhibit two interesting characteristics, as shown in Fig. \ref{network}. On one hand, agents demonstrate homogeneous connectivity where they prioritize connections with like-minded agents who share similar stances, aggregating distinctly into local clusters. On the other hand, agents with neutral attitudes act as structural bridges, more evenly dispersed between different clusters, thereby facilitating inter-group connectivity. 
These findings provide insights into possible mitigation strategies of echo chambers and attitude polarization.

\section{Conclusion}


In this work, we propose $DynamiX$, a large-scale social network simulator  designed to explore dynamic social networks.
$DynamiX$ reflects the dynamic switching of agents roles and dynamic evolution of social relationships, ensuring high-fidelity and high-precision in large-scale social simulations. 
Compared to existing static simulators, $DynamiX$ not only reveals the intrinsic significance of dynamic social networks in attitude dynamics and collective interactions, but also provides a novel perspective for the cultivation of opinion leaders. 
Together, these contributions underscore the potential of $DynamiX$ as a testbed for investigating social dynamics, collective phenomena, follower growth, and large-scale simulations across the human sciences. 

While $DynamiX$ offers valuable insights into social dynamics, it simplifies behaviors space and memory mechanisms of agents, and relies on generated content with inherent biases, highlighting the need for further advancements to better capture the complexity of human interactions. Additionally, the absence of  real-time, real-world information prevents $DynamiX$ from aligning with intense real-world fluctuations, resulting in a smooth dynamics. Future work will focus on integrating external tools, multimodal information, and other virtual-real alignment methods to enhance the realism, adaptability, and scalability of $DynamiX$, ultimately bridging the gap between simulated and real-world social environments.

\section*{ETHICAL STATEMENT}

The development and application of $DynamiX$ require careful ethical consideration. To prevent misuse for manipulating public opinion, transparency and strict supervision are essential. They involve standardizing usage scope, clarifying permissions, and defining ethical boundaries to avoid undue influence on public attitudes. Additionally, the privacy and confidentiality of real-world data used in simulations are critical. We are committed to ensuring user privacy, data reliability, and preventing unauthorized access and disclosure of sensitive information. Lastly, though $DynamiX$ serves the public interest, there exists a risk of disseminating harmful or biased content, particularly on controversial topics. To mitigate this, robust verification and control mechanisms will be implemented to prevent the spread of harmful information. By adhering to stringent standards and safeguards, $DynamiX$ will be used solely for legitimate purposes, avoiding any form of information manipulation or illegal activities.

\bibliography{ustc}
\bibliographystyle{IEEEtran}

\end{document}